\documentclass[11pt]{article}
\usepackage{amsmath,amssymb,multirow,jheppub,graphicx}
\usepackage{centernot}
\usepackage{color}
\allowdisplaybreaks[1]
\usepackage{tikz}
 \tikzset{node distance=2cm, auto}
\usepackage{bbm} 
\usepackage{youngtab}
\usepackage{slashed}

\usepackage[shortlabels]{enumitem}

\def\ZZ{\mathbb{Z}}

\def\tfrac#1#2{{\textstyle{\frac{#1}{#2}}}}

\def\bar{\overline}

\def\b{{\beta}}

\def\t{{\tau}}

\def\Z{{\mathbb Z}}
\def\R{{\mathbb R}}
\def\coeff#1#2{{\textstyle {\frac {#1}{#2}}}}

\def\half{\coeff 12}

\usepackage{verbatim}   
\usepackage[all]{xy}
\usepackage{bm}  
\def\Dslash{{\rlap{\raise 1pt \hbox{$\>/$}}D}}
\def\Pslash{{\rlap{\raise  1pt \hbox{$\>/$}}\,\partial}}

\newcommand{\be}{\begin{equation}}      
\newcommand{\ee}{\end{equation}}      
\newcommand{\bea}{\begin{eqnarray}}      
\newcommand{\eea}{\end{eqnarray}}

\newcommand{\im}{\mathrm{i}}

\newcommand{\formabs}[1]{\left\Arrowvert #1 \right\Arrowvert}

\title{Four-fermion deformations of the massless Schwinger model and confinement}
\author[1]{Aleksey Cherman,}
\author[1]{Theodore Jacobson,}
\emailAdd{acherman@umn.edu}
\emailAdd{jaco2585@umn.edu}
\author[2]{Mikhail Shifman,}
\emailAdd{shifman@umn.edu}
\author[3]{Mithat  \"Unsal,}
\emailAdd{unsal.mithat@gmail.com}
\author[2,4]{Arkady Vainshtein}
\emailAdd{vainshte@umn.edu}
\affiliation[1]{School of Physics and Astronomy, University of Minnesota, Minneapolis, MN 55401, USA}
\affiliation[2]{Fine Theoretical Physics Institute, School of Physics and Astronomy, University of Minnesota, Minneapolis, MN 55401, USA}
\affiliation[3]{Department of Physics, North Carolina State University, Raleigh, NC 27607, USA}
\affiliation[4]{KITP, Santa Barbara, USA}

\abstract{We consider the massless charge-$N$ Schwinger model and its deformation with
two four-fermion operators. Without the deformations, this model exhibits chiral symmetry
breaking without  confinement. It is usually asserted that  the massless
Schwinger model is always deconfined and a string tension  emerges only
when a mass for the fermion field is turned on.    We show that in the presence of these four-fermion operators, the
massless theory can in fact
confine.  One of the four-fermion deformations is chirally neutral, and
is a marginal deformation. The other operator can be relevant or irrelevant, and respects a  $\Z_2$ subgroup of
chiral symmetry for even $N$, hence forbidding a mass term. When it is relevant, even the exactly
massless theory exhibits  both confinement and spontaneous chiral symmetry
breaking.   The construction is analogous to  QCD(adj) in 2d. While the
theory without  four-fermion deformations is deconfined, the theory with these
deformations is generically in a confining phase. We study the model on $\mathbb{R}^2$ using bosonization, and also analyze the mechanism of confinement on $\mathbb{R}\times S^1$, where we find that confinement is driven by fractional instantons.
}

\begin{document}
\maketitle

\section{Introduction}
The Schwinger model---U(1) gauge theory coupled to a Dirac fermion in two
spacetime dimensions---is a famous playground for the exploration of ideas 
about quark confinement and chiral symmetry breaking~\cite{PhysRev.128.2425,Coleman:1975pw,Coleman:1976uz}.  These topics are
notoriously difficult to study in non-abelian gauge theories in four spacetime
dimensions, so it is very helpful to be able to explore them in the
calculable setting of the Schwinger model.

Historically, the Schwinger model was introduced as $U(1)$ quantum
electrodynamics with a single unit-charge Dirac fermion.\footnote{\,The
seminal works from the 1970s, e.g. \cite{Coleman:1975pw,Coleman:1976uz},
considered the response of the theory to the introduction of test
charges with irrational charge, so technically these works took the
group manifold to be a real line $\mathbb{R}$.} Here we will study some
generalizations of this theory with the aim of making a better toy model
for 4d gauge theory. Our starting point will be the charge-$N$ Schwinger
model, which was studied recently in e.g.
Refs.~\cite{Anber:2018jdf,Anber:2018xek,Armoni:2018bga,Misumi:2019dwq,Komargodski:2020mxz,Cherman:2020cvw,Cherman:2021nox}, see also \cite{Hansson:1994ep} for an early analysis.
This model has the Euclidean action
\begin{align}
  S_{\rm standard} = \int d^2 x\,\left( \frac{1}{4e^2} f_{\mu\nu}^2 
  + \bar{\psi} \left[ \gamma^{\mu} (\partial_{\mu} + iN a_{\mu}) \right] \psi \right)
 +  m_\psi \bar{\psi}_L\psi_R + \textrm{h.c.}\,.
  \label{eq:standard_Schwinger}
\end{align}
Here $a_{\mu}$ is a $U(1)$ gauge field, $\psi$ is a Dirac fermion field with chiral components $\psi_L,\psi_R$, the
integer $N$ is the charge of the fermion, $e$ has unit mass dimension, and Hermitian conjugation is defined by analytic continuation from Minkowski space. 
The
statement that $a_{\mu}$ is a $U(1)$ gauge field means that the gauge
transformation functions $\alpha(x)$ take values in $U(1)$, meaning that
there is $2\pi$ periodicity. Physically, one could interpret this
periodicity as arising from gauge transformations of a very heavy
unit-charge test fermion field $\psi_t$, 
\begin{align*}
  \psi_t &\to e^{i \alpha} \psi_t,
\end{align*}
while the gauge
transformations of the other fields are 
\begin{align*}
  a_{\mu} &\to a_{\mu} - \partial_{\mu} \alpha, \\[1mm]
  \psi &\to e^{i N \alpha} \psi. 
\end{align*}
Gauge invariance also implies that $\int_{M_2} f \in 2\pi \mathbb{Z}$
where $M_2$ is any closed smooth 2-manifold and $f =da= \frac{1}{2}f_{\mu \nu}dx^{\mu} \wedge dx^{\nu}$ is the field strength 2-form.

We also add an explicit topological $\theta$ term 
\begin{align}
  S_{\theta} = \frac{i \theta}{2\pi} \int_{M_2} da 
  \label{top}
\end{align}
to the action, and assume that $m_\psi \ge 0$.    The coefficient $\theta$ is $2\pi$ periodic.  However,
we will mostly focus on the physics at $\theta = 0$, where the theory
has a $\mathbb{Z}_2$ parity symmetry.

As defined above, the Schwinger model has a $\mathbb{Z}_N$ 1-form
symmetry~\cite{Gaiotto:2014kfa} for any value of $m$.  This symmetry is
generated by a collection of $N$ local topological operators $U_n(x)$
and has the effect of multiplying Wilson loops by $\mathbb{Z}_N$ phases:
\begin{align}
  \langle U_{n}(x) \, e^{i q \int_C a_{\mu} dx^{\mu} } \rangle
  = \exp\left(\frac{2\pi i n q }{N}\ell(C, x)\right)   \langle e^{i q \int_C a_{\mu} dx^{\mu} } \rangle 
\end{align}
where $\ell(C,x)$ is the linking number of $C$ and $x$. The internal
global symmetry of the $m \neq 0$ Schwinger model---which is just the
$\mathbb{Z}_N$ 1-form symmetry---coincides with the global symmetry of
pure 4d  $SU(N)$ Yang-Mills theory. The existence of the 1-form
$\mathbb{Z}_N$  symmetry means that charge confinement is a
sharply-defined concept in the Schwinger model when $N>1$.  The same is
true in 4d $SU(N)$ pure YM theory.\footnote{In discussions of 4d YM
theory it is common to call its $\mathbb{Z}_N$ 1-form symmetry ``center
symmetry''~\cite{Polyakov:1976fu,Gross:1980br}.  This has some historical
justification because the center subgroup of $SU(N)$ is $\mathbb{Z}_N$,
which happens to be the same as the $1$-form symmetry group so long as
all matter fields are in representations of $N$-ality zero. We won't use
this language here because the addition of matter reduces the 1-form symmetry to a discrete subgroup, while the center subgroup of the $U(1)$ gauge group is $U(1)$. }

When $m_\psi=0$, at the classical level the Schwinger model has a $U(1)$ axial
symmetry. As usual, the ABJ anomaly means that the chiral symmetry in the
quantum theory is reduced, and is generated by 
\begin{align}
  \psi(x)  \to e^{2\pi i \gamma_5/(2N)} \psi(x) \,.
\end{align}  
The faithfully-acting symmetry is $\mathbb{Z}_N$, and acts as
$\bar{\psi}_L \psi_R \to e^{ 2\pi i/N}\bar{\psi}_L \psi_R$.   The
existence of the discrete chiral symmetry means that it is meaningful to
discuss spontaneous chiral symmetry breaking, just as in e.g 4d $SU(N)$
$\mathcal{N}=1$ super-YM theory.  The $\mathbb{Z}_N$ $0$-form chiral
symmetry and the $\mathbb{Z}_N$ $1$-form symmetry have a mixed 't Hooft
anomaly~\cite{Anber:2018jdf}. Indeed,  the internal global symmetries and
anomalies of the massless charge-$N$ Schwinger model coincide with the
internal bosonic global symmetries and anomalies of 4d $SU(N)$
$\mathcal{N}=1$ super-YM theory.

From the perspective of the first paragraph of this introduction, it
would be nice if the dynamics of 4d $SU(N)$ gauge theories and the
charge-$N$ Schwinger model looked similar.  Unfortunately,  the behavior
of the 4d and 2d theories is very different!  When $m_\psi = 0$, the
good news is that the $\mathbb{Z}_N$ chiral symmetry is spontaneously
broken, just as in 4d $\mathcal{N}=1$ SYM.  The bad news for the
comparison with 4d gauge theory is that the $\mathbb{Z}_N$ 1-form
symmetry of the Schwinger model with $m_{\psi} = 0$ is spontaneously
broken, and the expectation values of large `fundamental' Wilson loops
have a perimeter-law behavior
\begin{align}
  \langle e^{i \int_C a}\rangle \sim e^{-\mu P(C)} \,,
\end{align}
where $C$ is e.g. a circular contour with perimeter $P(C)$, $\mu$ is a UV scale (of order the mass of a heavy test particle),
and $a = a_{\mu} dx^{\mu}$.  
Confinement appears (and the 1-form symmetry is restored) when $m_\psi
\neq 0$, but the string tension scales as $T \sim m_\psi e$ for $m_\psi
\ll e$.   In 4d $\mathcal{N}=1$ SYM, in contrast, the $1$-form
$\mathbb{Z}_N$ symmetry is not spontaneously broken, and large Wilson
loops obey an area law.

There are three basic ways
to understand the behavior of Wilson loops in the Schwinger model:
\begin{enumerate}[(a)]
  \item  Solve the charge-$N$ Schwinger model exactly on $\mathbb{R}^2$ using 
  bosonization and compute the
  relevant expectation values.  This has the virtue of using direct and
  relatively elementary arguments.   
  \item Relate deconfinement to the existence of a mixed 't Hooft anomaly
  between the $\mathbb{Z}_N$ $1$-form symmetry and the $\mathbb{Z}_N$
  $0$-form chiral symmetry. This approach has the advantage that it uses
  only basic symmetry principles, and so it generalizes to theories which
  are not exactly solvable. 
  \item Solve the model on $\mathbb{R}\times S^1$ with small $S^1$ and extrapolate the phase structure to $\mathbb{R}^2$.  Due to the 't Hooft anomaly, the quantum-mechanical EFT has $N$ degenerate ground states.  One can take any linear combination of them to be a ground state.  The two most interesting choices result in either the chiral condensate being non-zero  while the Polyakov loop has zero expectation value, or vice versa.  Only the former choice extrapolates nicely to $\mathbb{R}^2$.  The issue with the other choice is that in the large $S^1$ limit the Polyakov loop disappears as an observable. We should emphasize that in 2d gauge theories, the vanishing of the Polyakov loop expectation value on a cylinder $S^1$ is not sufficient to conclude that the $1$-form symmetry is unbroken on $\mathbb{R}^2$.
\end{enumerate}
All of these approaches have been discussed in the
literature~\cite{Anber:2018jdf,Anber:2018xek,Armoni:2018bga,Misumi:2019dwq,Komargodski:2020mxz,Cherman:2020cvw,Cherman:2021nox}.

The fact that charge-$q$ Wilson loops are deconfined for all $q \in
\mathbb{Z}$ in the charge-$N$ Schwinger model sharply contrasts with the
expected behavior in 4d $SU(N)$ gauge theory with adjoint fermions. The
latter theory is expected to confine fundamental test charges even when
the mass of the dynamical fermion goes to zero, at least as it is
outside of the conformal window.  Here we discuss a modification of the
charge-$N$ Schwinger model which brings its dynamics much closer to
the dynamics of 4d gauge theories.   We will mostly focus on the
Schwinger model with even $N$ for reasons that will become clear
shortly. 

We now explain the basic idea of this paper.  First, we recall that at high energies the standard Schwinger model \eqref{eq:standard_Schwinger} approaches a free-field CFT fixed point, which can be described as a free massless Dirac fermion.  We then note that this CFT contains a unique exactly marginal operator
\begin{align}
    \mathcal{O}_{\rm jj}
&= j_{\mu} j^{\mu} = \bar{\psi} \gamma_{\mu} \psi\, \bar{\psi} \gamma^{\mu} \psi = - 4\bar{\psi}_R \psi_L \bar{\psi}_L \psi_R \,.
\label{eq:Ojj}
\end{align}
This `Thirring model' operator is exactly marginal, $\Delta_{\rm jj} = 2$, and neutral under all symmetries of the model.  Therefore we are free to add it to the action with a dimensionless coefficient $g \in \R$. This yields a generalization of the Schwinger model which we will call the Schwinger-Thirring (ST) model:
\begin{align}
S_{\rm ST} = S_{\rm standard} + g \int d^2x\, \mathcal{O}_{jj}
\label{eq:ST_action}
\end{align}
Since \eqref{eq:ST_action} is an interacting theory, in general one might expect a
dimensionless parameter like $g$ to run with the RG scale.  However,
when $m_{\psi} = 0$ it is known that in fact $g$ remains an exactly marginal
parameter even after we take into account the gauge interaction~\cite{Thirring:1958in,Johnson:1961cs,Mueller:1971mar,Gomes:1972yb}. This is easiest to see
using bosonization, as we review below, but it can also be deduced directly in the fermionic variables, see Appendix~\ref{sec:scaling_dim}. There is a minimum value of $g$, $g_{*} = -\pi/2$,
below which some operator scaling dimensions become negative, and the theory ceases to be unitary.  We will assume that $g > g_{*}$. 
Turning on the perturbation by $\mathcal{O}_{\rm jj}$ perturbation does not affect the
symmetries and anomalies of the massless Schwinger model.  As a result, the
massless Schwinger-Thirring model remains in a deconfined phase with a
finite mass gap and spontaneous chiral symmetry breaking for $g \in
(g_*,\infty)$, just like the original charge $N$ Schwinger model.

We should emphasize that the high energy behavior of $S_{\rm ST}$ is \emph{not} the same as that of the original Schwinger model, although it is continuously connected to it.  Rather than approaching a free-field CFT fixed point, it approaches an \emph{interacting} CFT fixed point at high energies.  In this UV CFT fixed point the scaling dimensions of operators do not coincide with their `engineering' dimensions.

The lowest-dimension four-fermion operator which is invariant under parity but not $\mathbb{Z}_N$ chiral symmetry is 
\begin{align}
\mathcal{O}_{\chi} 
&= \bar{\psi}_L \psi_R (D_{\mu}\bar{\psi}_L) (D^{\mu} \psi_R ) \,.
\label{eq:Ochi}
\end{align}
At the free-field fixed point (that is, at high energies and with $g=0$), the scaling dimension of $\mathcal{O}_{\chi} $ coincides with its engineering dimension, which is $4$. So when $g=0$, this operator is RG irrelevant. But the scaling dimension $\Delta_{\chi}$ of $\mathcal{O}_{\chi}$ depends on $g$, and we will show that $\Delta_{\chi}$ decreases monotonically as $g$ is increased.  Specifically, bosonization implies that
\begin{align}
  \Delta = \frac{4}{1+2g/\pi }\,.
  \label{eq:Ochi_dim}
\end{align}
The scaling dimension of $O_{\rm \chi}$ diverges as $g \to g_* = -\pi/2$ from
above, but it becomes relevant when $g >\pi/2$.

As a result, there is a critical value of $g$ ($g = \pi/2$) at which $\mathcal{O}_{\chi}$ becomes marginal at the UV fixed point, and as $g$ is increased further, the $\mathcal{O}_{\chi}$ operator becomes relevant at the UV fixed point.  
At the same time, we note that the operator $\mathcal{O}_{\chi}$ is the
lowest-dimension operator with charge $2$ under $\mathbb{Z}_N$ chiral
symmetry, and it is invariant under all other symmetries.  In the rest of this paper, we will discuss what happens to the low-energy physics once we add the $\mathcal{O}_{\chi}$ operator to the UV action of the ST model as a perturbation:\footnote{An
inspirational brief discussion of a very similar deformation was given in
Ref.~\cite{Komargodski:2020mxz}.  However, the form of the deformation
in the fermionic variables given in Ref.~\cite{Komargodski:2020mxz} is
the same as our $\mathcal{O}_{\chi}$ without derivatives, which vanishes
identically due to fermi statistics.  The form of the deformation in the
bosonic variables in Ref.~\cite{Komargodski:2020mxz} coincides with our
bosonized expressions, but the consequences of the technical irrelevance
of this operator were not highlighted.}
\begin{align}
  S = S_{\rm ST} 
  + \Lambda^{2-\Delta_{\chi}}\int d^2x\,   (\mathcal{O}_{\chi}+ \mathcal{O}_{\chi}^\dagger) \,.
  \label{eq:our_model}
\end{align}
The parameter $\Lambda$ is a new parameter with unit mass dimension.  Its power is fixed from the scaling dimension of $\mathcal{O}_{\chi}$ at the UV fixed point.  Whether one should think of $\Lambda$ as an IR or a UV energy scale depends on $\Delta_{\chi}$, and as we have already said $\Delta_{\chi}$ depends on the marginal parameter $g$.  If $\Delta_{\chi}>2$, then $\Lambda$ is a UV scale: the
model defined by Eq.~\eqref{eq:our_model} needs a UV completion at the scale $\Lambda$. In this case we will get a physically-interesting model if $e \ll \Lambda$.  If $\Delta_{\chi}<2$, then $\Lambda$ is an IR mass scale in the same sense as the fermion mass parameter is an `IR scale' of a free-fermion theory, and there is no a priori constraint on the ratio $\Lambda/e$.  Nevertheless, we will see that the model is weakly coupled (in the bosonized duality frame) when $\Lambda/e \ll 1$, and so that is the regime we will focus on.

\begin{figure}[t]
\begin{center}
\includegraphics[width = 0.9\textwidth]{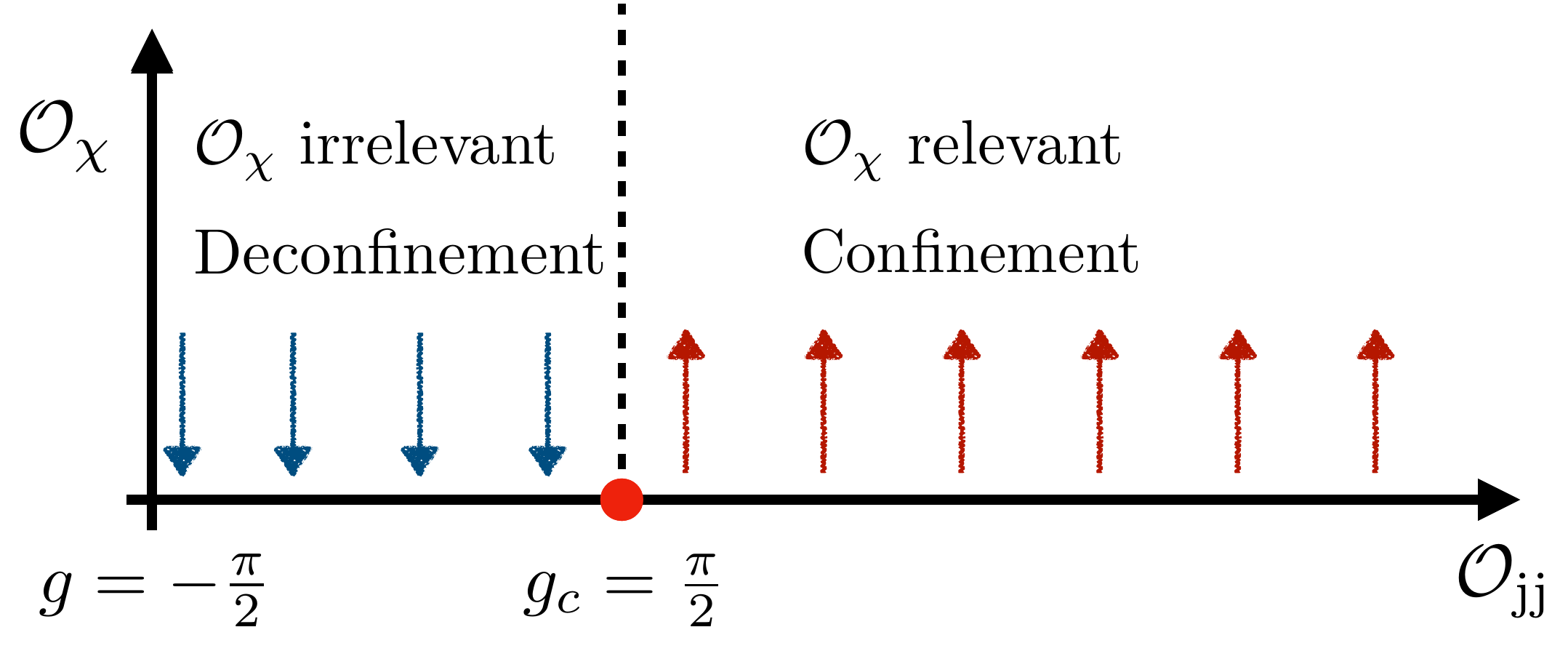}
\caption{Behaviour of the massless Schwinger model in the presence of two four-fermion operators  
${\cal O}_{jj}$ and ${\cal O}_{\chi}$ for  $N$-even. ${\cal O}_{jj}$ is a marginal deformation. 
When $g>g_c$, ${\cal O}_{\chi}$ becomes relevant and the massless theory becomes  confining. 
The behaviour of the four-fermion-deformed Schwinger model is similar in this respect to four-fermion-deformed QCD(adj)$_2$. 
 }
\label{fig:confine}
\end{center}
\end{figure}

To get a feeling for the physics of the four-fermion deformed model \eqref{eq:our_model} let us first suppose that $\Delta_{\chi} >2$.  This would be the case if e.g. we set $g=0$. Then chiral symmetry is
explicitly broken at the UV scale $\Lambda$. For even $N$ it is broken to
 a $\mathbb{Z}_2$ subgroup, while for odd $N$ it is broken
completely.   One might therefore expect that the model becomes confining for odd $N$.
However, this is not quite correct.  At long distances $\ell \gg 1/e \gg
1/\Lambda$, the deformation by $\mathcal{O}_{\chi}$ is irrelevant in the
RG sense, and at distances large compared to $1/\Lambda$ there can be an
emergent $\mathbb{Z}_N$ chiral symmetry.\footnote{When the $\mathcal{O}_{\chi}$ operator is irrelevant, we expect it to behave similarly to the Wilson term in the Wilson fermion action in 4d lattice gauge theory.  In 4d, the Wilson term is an irrelevant dimension $5$ operator, schematically $\bar{\psi} D_{\mu}D^{\mu} \psi$, which breaks chiral symmetry as well as the degeneracy between the 16 `doubler' fermion modes.   If the bare mass term is set to zero, then a lattice-scale mass term is induced by RG flow, and there is no emergent chiral symmetry in the infrared:  the Wilson term is dangerously irrelevant.   But one can get an emergent chiral symmetry in the infrared by tuning the bare quark mass term.   We expect the same to be true for the $\mathcal{O}_{\chi}$ operator when $N$ is odd, but leave a detailed exploration of this feature of the model to future work. 
\label{footnote:Wilson}}  Indeed, in this case the string tension
induced by the $\Lambda$ perturbation is proportional to e.g. 
$e^4/\Lambda^2$ when $g = 0$, 
and so unit test charges separated by a distance $L$
satisfying $1/e \ll L \ll \Lambda^2/e^3$ will not feel a linear potential.
In this sense the ST model with a deformation by
$\mathcal{O}_{\chi}$ with $\Delta_{\chi}>2$ is no more confining than $U(1)$ QED in three
spacetime dimensions, interpreted as a lattice gauge theory on a square Euclidean lattice with a Wilson action.   In that model, at finite lattice spacing $a$
there are finite-action monopole instantons which induce a finite string
tension~\cite{Polyakov:1976fu}.  But the monopole-instanton action
diverges in the continuum limit $a \to 0$, so the string tension also
goes to zero in the continuum limit.\footnote{Recently it was understood that there are other lattice actions which flow to $U(1)$ QED in three spacetime dimensions without any finite-action monopole-instantons even at finite lattice spacing~\cite{Sulejmanpasic:2019ytl}, see also e.g.~\cite{Gorantla:2021svj}.}

We will show that when $g \ge \pi/2$, our deformed Schwinger model \eqref{eq:our_model} confines fundamental (that is, $q = \pm 1$) test charges when $N>2$.  When
$N$ is even the $\mathbb{Z}_N$ 1-form symmetry is spontaneously broken
to $\mathbb{Z}_{N/2}$,  so test charges with $q = N/2 \textrm{ mod } N$
are deconfined, while others are confined.  Also, when $N$ is even, the model
has a $\mathbb{Z}_2$ chiral symmetry, the fermion mass term is
forbidden, and one can think of   \eqref{eq:our_model} as a variant of the massless charge-$N$ Schwinger
model with \emph{confinement} for $N>2$.  When $N$ is odd and larger than $1$,
chiral symmetry is completely broken, and the $\mathbb{Z}_N$ 1-form
symmetry is not spontaneously broken at all.  The mass term can be
generated by fluctuations.  We will show how these features arise using
bosonization on $\mathbb{R}^2$, as well as by an analysis on $\mathbb{R}
\times S^1$ when $S^1$ is small.  These analyses have complementary
strengths, and combining them yields some interesting insights into the
nature of the confinement mechanism in this model.

The behavior of the four-fermion-deformed charge-$N$ Schwinger model is much closer to the expected behavior of QCD-like 4d gauge theories.  The four-fermion-deformed Schwinger model is also a nice toy model for the behavior of 2d $SU(N)$ adjoint QCD.  Adjoint QCD in 2d has only a $\mathbb{Z}_2$ chiral symmetry when the quark mass is set to zero, and is also known to have two interesting four-fermion deformations~\cite{Cherman:2019hbq} consistent with chiral symmetry.  When these deformations are tuned to zero, 2d adjoint QCD deconfines on $\mathbb{R}^2$ due to a mixed \mbox{'t Hooft} anomaly between its $\Z_N$ 1-form symmetry and an exotic non-invertible symmetry~\cite{Komargodski:2020mxz}.  However, once the four-fermion deformations are turned on, at generic points in its parameter space 2d adjoint QCD confines~\cite{Cherman:2019hbq}.

\section{Confinement from elementary considerations}
\label{sec:confinement_elementary}
\subsection{Confinement in the standard charge $N$ Schwinger model}

As discussed in the introduction, the  massless charge-$N$ Schwinger model   has
a $\Z_N^{(1)}$ 1-form symmetry and  
$\Z_N^{(0)}$ 0-form  chiral symmetry.   It is often asserted that  when the
fermions are massless, the theory does not confine integer test charges, while
with massive fermions, it does confine integer test charges.    The common
argument for this involves considering the topological $\theta$ parameter of
$U(1)$ gauge theory, which enters the Euclidean action through
\begin{align}
  S_{\theta} = \frac{i \theta}{2\pi} \int_{M_2} da \,.
\end{align}
Coleman observed that changing
$\theta$ by $2\pi$ corresponds to inserting a particle of charge $\pm 1$ at $x =
\pm \infty$.  This means that the $k$-string tension can be written as 
\begin{align}
T_k (\theta) =  \mathcal{E}(\theta +2\pi k) -    \mathcal{E}(\theta)\,, 
\label{ten1}
\end{align} 
where   $\mathcal{E}(\theta)$ is the vacuum energy density as a function of $\theta$.

Of course, when $m_\psi =0$, there is no $\theta$ dependence in vacuum energy,
because a chiral rotation can remove the $\theta$ term from the action.  This
immediately implies that the massless theory does not confine integer test
charges. Once a mass term for fermions is added,  the $\theta$ term can no
longer be removed by chiral rotations:  a transformation that would remove the
topological term from the action reintroduces it  in the mass term as $m_\psi
\overline \psi_L \psi_R \rightarrow  m_\psi   e^{ i \theta/N} \overline \psi_L
\psi_R$. As a result, when $m_{\psi} \neq 0$, the degeneracy between the   $N$
chirally broken vacua  is lifted. When $m_{\psi}$ is small, the $\theta$ dependence
of the vacuum energy density  emerges as $ \mathcal{E}_k (\theta) = - m_\psi
\langle  \overline \psi_L  \psi_R \rangle + {\rm c.c} $ where   the chiral
condensate is given by (see e.g.~\cite{Sachs:1991en}):
\begin{align}
  \langle  \overline \psi_L  \psi_R \rangle  = \frac{m_\gamma \, e^{\gamma}}{4 \pi} e^{ i \frac{\theta+ 2 \pi k}{N}}, \qquad k=1, \ldots, N 
\end{align}
and $m_\gamma = {Ne}/\sqrt{\pi} $ is the mass gap in the theory.\footnote{We use the same symbol $e$ for the the base of the natural logarithm and the gauge coupling, and hope that readers can distinguish them from context.} The string
tension for a charge-$k$ probe in the presence of the theta angle can be written
as
\begin{align}
T_k(\theta)  = - m_\psi \mu \frac{N e \,}{2 \pi^{3/2} }  \left[   \cos  \left(  \frac{ \theta + 2\pi k}{N}  \right) -  \cos  \left(  \frac{ \theta }{N}  \right)  \right] + O(m_\psi^2)\,.
\end{align} 
For $\theta=0$, this expression can be simplified into 
\begin{align}
T_k(\theta=0)  =  m_\psi \mu \frac{N e\,}{ \pi^{3/2} }    \sin^2  \left(  \frac{ \pi k}{N}  \right) +  O(m_\psi^2)\,.
\label{tension-strong}
\end{align} 
In these formulas $\mu$ is a renormalization scale, which of course would cancel in appropriate ratios of dimensionful physical quantities.  Clearly, there is a finite tension, and hence confinement,  for  charges  $k
\neq 0  \; ({\rm mod} \;  N)$, and charges that are multiples of $N$ are
screened: 
\begin{align} 
{\rm mass \; deformation:} \qquad \langle W_k(C)  \rangle  = \left\{ 
\begin{array}{ll} 
e^{- T_kA(C) },  & \qquad  k \neq 0   \qquad ({\rm mod}\;  N) \\
e^{-M P(C)  }, &  \qquad k = 0  \qquad  ({\rm mod} \; N) 
\end{array} \right.\,.
\end{align}
where $A(C)$ is the area of the disk-like region enclosed by the curve $C$,
$P(C)$ is the perimeter of $C$,  $M$ is a non-universal mass scale, and we
have assumed that $A(C)$  is large compared to the microscopic scales of the
theory, while at the same time it is small compared to the size of the spacetime
manifold.

\subsection{Confinement in the four-fermion deformed charge-$N$ Schwinger model}
\label{sec:elementary_confinement}
The discussion above may lead one to think that it is necessary to have  massive
 fermions to achieve confinement in the Schwinger model. However, this is not
 true. All we need is for the vacuum energy density to have non-trivial
 $\theta$-dependence. In fact,  even when the fermions are exactly massless and
 a  chiral symmetry protects a mass term from being generated, the gauge interactions in the Schwinger
 model may lead to confinement of fundamental test charges.  For this to be the case, what we need is a
 deformation which is chirally charged, so that with its inclusion, the $\theta$
 term cannot be removed, while at the same time the deformation preserves a
 non-trivial  subgroup of the chiral symmetry  so that the mass term is still
 forbidden.  Finally, we want the deformation operator to be marginal or
 relevant, so that its effects survive at long distances.  We will defer a
 discussion of this last point to the next section, and focus on the first point
 here. 
 
Consider the chirally-charged four-fermion operator in Eq.~\eqref{eq:Ochi}. The ABJ
anomaly reduces $U(1)_A$ down to $\Z_{2N}$, but the $\Z_2$ part of this
transformation is part of the gauge redundancy.  Therefore the faithful symmetry
is only $\Z_{N}$, as explained earlier, and the four-fermion deformation breaks the
anomaly-free faithfully-acting $\Z_{N}$ chiral symmetry down to $\Z_2$ for $N$
even and breaks it completely for $N$ odd. Therefore, for $N$ even, a mass term
cannot be generated when the deformation is turned on. However, if $N$ is odd, a
mass term can be generated non-perturbatively.  A heuristic argument for this
goes as follows. Pure $U(1)$ gauge theory on a torus has instantons with integer
topological charge $\frac{1}{2\pi} \int_{T^2} f \in \mathbb{Z}$.  If we add a
massless charge $N$ Dirac fermion, a charge $1$ instanton has $2N$ zero modes.
When $N$ is odd we can soak up its fermion zero modes $\frac{N-1}{2}$ times by
using the four-fermion operators and generate a fermion mass term from a sum over a
dilute gas of instantons.  However, this picture is only heuristic because when $T^2$
is large compared to the gauge coupling $e$, instantons are not localized, so a
dilute instanton gas sum does not make sense.  In Section
\ref{sec:circle_physics} we will discuss a regime where a semiclassical
calculation involving finite-action field configuration does make sense, and
make these remarks more precise.

\begin{figure}[t]
\vspace{-0.5cm}
\begin{center}
\includegraphics[width = 0.9\textwidth]{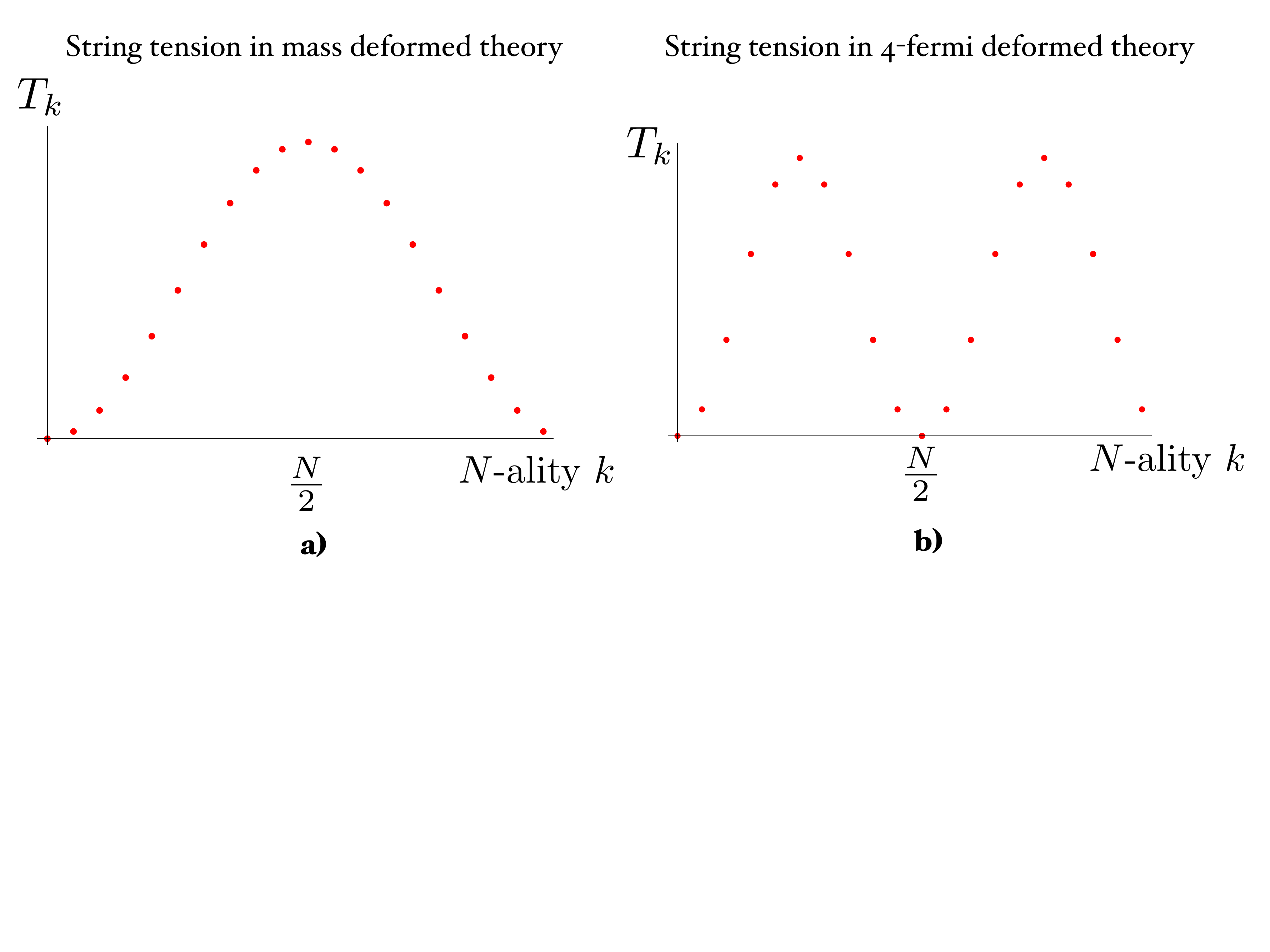}
\vspace{-4.5cm}
\caption{String tensions in the charge-$N$ Schwinger model for even $N$ as a
function of $N$-ality in a) the mass-deformed theory and b) the theory deformed
by the two four-fermion operators. The first case exhibits a single hump structure,
and the string tension is maximal at $k = \frac{N}{2}$, and there is generically
a two-fold degeneracy of tensions. In the second case, the string tension
vanishes at $k = \frac{N}{2}$,  there is a double-hump structure, and the spectrum of string
tensions is generically 4-fold degenerate. }
\label{fig:Tensions}
\end{center}
\end{figure}

If we do a chiral rotation to remove the topological term \eqref{top} in the
action, we reintroduce $\theta$ in the chirally-charged four-fermion operator as $O_{\rm \chi}
\mapsto e^{2 i \theta/N} \mathcal{O}_{\chi}$. Therefore, even in the absence of massless
fermions, the vacuum energy density depends on $\theta$.
Following the same steps as in the undeformed theory, we find that the string
tension  for a charge $k$ probe is 
\begin{align}
T_k  \sim - \Lambda^{2-\Delta_\chi} 
\left[   \cos    \left(  \frac{ 2 (\theta + 2\pi k)}{N}  \right) -  \cos  \left(  \frac{ 2 \theta }{N}  \right)  \right] 
+ \mathcal{O}(\Lambda^{2(2-\Delta_\chi)} ) \,,
\end{align} 
where $\Delta_\chi$ is the scaling dimension of $O_{\rm \chi}$. If $\Delta_\chi > 2$,
$\Lambda$ is a UV scale, and $T_k$ vanishes for all $k$ as we take $\Lambda/e \gg 1$.  If
$\Delta_\chi < 2$, then $\Lambda$ is an IR scale.  The expression above assumes that $\Lambda/e \ll 1$ when $\Delta_\chi <2$.
For $\theta=0$, this expression can be simplified into 
\begin{align}
T_k  \sim  \Lambda^{2-\Delta_\chi}    \sin^2  \left(  \frac{ 2 \pi k}{N}  \right)  + \mathcal{O}(\Lambda^{2(2-\Delta_\chi)} ) \,.
\label{tension-strong-4fermi}
\end{align} 
Note that for $N$ even, the string tension vanishes for charges $k= 0, N/2 \;
({\rm mod}\;  N)$, and is non-vanishing otherwise. For $N$ odd,  string tensions
except for $k= 0 \; ({\rm mod}\;  N)$ are non-zero.    The basic fate of confinement is illustrated by the sketch in Fig.~\ref{fig:confine}.  An  interesting feature in
both cases  is the double-hump structure of the tensions as a function of $k$.
For example, for odd $N$,  the minimal tension is not $T_1 = T_{N-1}$,  but $
T_{(N-1)/2} = T_{(N+1)/2}$.  This is illustrated in Fig.~\ref{fig:Tensions}\,.

To summarize, in our version of the massless Schwinger model defined by
\eqref{eq:our_model} with $g > \pi/2$, large Wilson loops have the
following behavior: 
\begin{align} 
\langle W_k(C)  \rangle  = \left\{ 
\begin{array}{ll} 
e^{-T_k A(C) },  & \qquad  k \neq 0, N/2   \qquad ({\rm mod}\;  N) \\
e^{-M P(C)  }, &  \qquad k = 0, N/2  \qquad  ({\rm mod} \; N) 
\end{array} \right.\,.
\end{align}
The probe charges $k= 0, N/2 \; ({\rm mod}\;  N)$ are screened and  the other
probe charges  are confined. The main distinction relative to the standard massive Schwinger
model is the fact that $k=N/2$ probe charge is confined in the massive model,
and is screened in the four-fermion deformed model. In the semi-classical domain, we
will see the microscopic difference between these two versions of confinement.

\section{Bosonization}

It is famously useful to treat the Schwinger model using bosonization, and in this section we describe the bosonized form of the model.  This will allow us to understand the interplay of the two four-fermion deformations of the model, and will be very useful both for the analysis of the dynamics on $\mathbb{R}^2$, as well as to understand some subtleties that arise in our analysis of the physics on $\mathbb{R} \times S^1$.

Bosonization amounts to a `change of variables' in the path integral
from the fermion field $\psi$ to a scalar field $\varphi$.  The scalar
is circle-valued, $\varphi(x)\equiv \varphi(x)+2\pi$,  so e.g. $d\varphi$ and
$e^{i k\varphi}$ with integer $k$ are good local operators, but
$\varphi$ itself is not.  To write down the bosonized theory, consider a
free massless fermion with gauged fermion parity.  It has two $U(1)$
global symmetries, the vector-like symmetry $U(1)_V$ and the axial
symmetry $U(1)_A$, with conserved current 1-forms $j_V, j_A$
respectively. The bosonic action corresponding to the free-fermion
theory is
\begin{align}
  S_{\varphi, \rm free}  = \int_{M_2} \frac{1}{8\pi} \formabs{d\varphi}^2 
  \label{eq:free_scalar}
\end{align}
where $\formabs{C}^2 \equiv C \wedge \star C$ for any differential form $C$.
The conserved currents of the bosonic and fermionic theories are related
via 
\begin{align}
  j_V \leftrightarrow -\frac{1}{2\pi} \star d \varphi \,, \quad j_A \leftrightarrow \frac{i}{4\pi} d \varphi\,.
  \label{eq:current_map}
\end{align}
Chiral symmetry acts on $\varphi$ via $e^{i\varphi} \to e^{2\pi i /N}
e^{i\varphi}$.  The bosonic operator corresponding to the fermion
bilinears is
\begin{align}
  \bar{\psi}_L(x) \psi_R(x) \leftrightarrow -\frac{\mu e^{\gamma}}{2\pi} \,e^{i \varphi(x)} 
\end{align}
and $\mu$ is the renormalization scale.  This scale appears on the bosonic side
of the mapping because the two-point function of $\varphi$ calculated from the
action \eqref{eq:free_scalar} has logarithmic UV/IR sensitivity, $\langle
\varphi(x) \varphi(0)\rangle - \langle \varphi(0)^2 \rangle \sim \log(x \mu)$. When the
renormalization scale is changed from $\mu$ to $\mu'$, exponentials of $\varphi$
transform as~\cite{PhysRevD.11.2088}
\begin{align}
 e^{  i k  \varphi} \to \left|\frac{\mu'}{\mu}\right|^{k^2} \!\!
e^{ i k  \varphi}
\end{align}
where on the left the renormalization scale of $ e^{  i k  \varphi}$ is $\mu$ while on the right it is $\mu'$. 

We will also need relations between the four-fermion operators
$\mathcal{O}_{jj},\; \mathcal{O}_{\chi}$ and bosonized
quantities.  First, note that
\begin{align}
  |j_V|^2 =\bar{\psi}\gamma_{\mu} \psi \bar{\psi}\gamma^{\mu} \psi \leftrightarrow \frac{1}{4\pi^2} |d\varphi|^2 \,.
\end{align}
This relation follows from the bosonic substitution \eqref{eq:current_map} for the current $(j_V)^{\mu} = \bar{\psi} \gamma^{\mu} \psi$\,.
It implies that the bosonic dual of the Schwinger-Thirring model
with action \eqref{eq:ST_action} is
\begin{align}
  S_{\textrm{bosonized ST}}  = \int_{M_2} \left[\frac{1}{2e^2} \formabs{d a}^2 + \frac{1}{2}R^2 \formabs{d\varphi}^2 
  - m \mu \cos(\varphi) -\frac{i N}{2\pi} d\varphi \wedge a \right]\,,
  \label{eq:bosonized_ST_action}
   \end{align}
where $m = e^\gamma m_\psi/2\pi$, and  
\begin{align}
  R^2 = \frac{1}{4\pi} \left(1 +\frac{2g}{\pi}\right)\,.
  \label{eq:radius}
\end{align}
The parameter $R$ gives the `radius' of the canonically-normalized scalar
$\tilde{\varphi}$ associated to $\varphi$, so that the periodicity of
$\tilde{\varphi}$ is $2\pi R$.  When $g = 0$, $R = 1/\sqrt{4\pi}$, so
the periodicity of $\tilde{\varphi}$ is $\pi^{1/2}$. 

Note when $m = 0$, the action \eqref{eq:bosonized_ST_action} is quadratic in the
fields, and so it is a free field theory in the
bosonic duality frame for any $R$ (that is, any $g$).  This makes it clear that $g$ does not run in the ST
model: it is an
exactly marginal parameter.  The energy is bounded from below so long as $g >
-\pi/2$.  When $g \neq 0$, the renormalization scale-change relation becomes
\begin{align}
e^{ i k  \varphi} \to \left|\frac{\mu'}{\mu}\right|^{k^2/4\pi R^2} 
  e^{i k  \varphi}
\label{eq:scheme_change_finite_g}
\end{align}
Equation
\eqref{eq:scheme_change_finite_g} implies that the scaling dimension of $e^{
i k \varphi}$ is 
\begin{align}
 \Delta_k \equiv \Delta[e^{ i k \varphi}] = \frac{k^2}{1+2g/\pi }\,,
\end{align}
so that the  bosonization rule for e.g. the fermion bilinear becomes
\begin{equation}
\bar{\psi}_L(x)\psi_R(x) \leftrightarrow -\frac{\mu^{\Delta_1}e^\gamma}{2\pi}e^{i\varphi(x)}\,.
\end{equation}
We can also state the bosonization rule for the
operator
\begin{align}
\mathcal{O}_{\chi}=  \bar{\psi}_L \psi_R (D_{\mu}\bar{\psi}_L) (D^{\mu} \psi_R )\,,
\end{align}
which is a scalar operator with chiral charge
$2$ and scaling dimension $4$ at $g=0$.  The only such operator in the bosonic description is
$e^{2i\varphi}$, so we conclude that
\begin{align}
  \mathcal{O}_{\chi}+\mathcal{O}_{\chi}^\dagger \;
  \longleftrightarrow\; c\mu^{\Delta_2}\, \cos(2\varphi)
  \label{eq:Ochi_bosonization}
\end{align}
where $c$ is an $\mathcal{O}(1)$ numerical
constant.\footnote{Our argument for Eq.~\eqref{eq:Ochi_bosonization} is based on matching symmetries and scaling dimensions.  We are not aware of a complete and explicit discussion of bosonization for this chirality-violating four-fermion operator in the  literature for the operator $\mathcal{O}_{\chi}$, although see Sec.\,5.6 of  Ref.~\cite{Fradkin:2013sab} for some interesting related discussion in a condensed-matter context.
}
This means that
\begin{align}
  \Lambda^{2-\Delta_2} \int d^2x \, (\mathcal{O}_{\chi}+\mathcal{O}_{\chi}^\dagger) \;
  \longleftrightarrow \; 
  c\mu^{\Delta_2}\, \Lambda^{2-\Delta_2} \int d^2x\,  \cos(2\varphi) \,.
\end{align}
Since we will be
working with the bosonized form of the theory from here onward, we will
absorb $c$ into the normalization of $\Lambda$, so it
will not appear in our
formulas. The bosonized action of the Schwinger model deformed by
$\mathcal{O}_{jj}$ and $\mathcal{O}_{\chi}$  with $m_{\psi} = 0$ is thus 
\begin{align}
  S = \int_{M_2} &\left[ 
   \frac{1}{2 }R^2 \formabs{d\varphi}^2 
   + \mu^{\Delta_2}\Lambda^{2-\Delta_2} \cos(2\varphi)  +\frac{1}{2e^2} \formabs{d a}^2  - \frac{i}{2\pi} N d\varphi \wedge  a \right]
  \label{eq:bosonized_with_gauge_field_mu}
\end{align}
In what follows we will often integrate the axion-like gauge field
coupling of $\varphi$ by
parts and drop the total derivative that appears in the process.  Appendix~\ref{sec:appendix}
contains a discussion of some interesting global subtleties of
axion interaction terms in 2d abelian gauge theories.  Handling these subtleties is important in some of our calculations
on $\mathbb{R} \times S^1$.  

The action in \eqref{eq:bosonized_with_gauge_field_mu} depends on the renormalization scale $\mu$.  When $\Delta_2 > 2$, the $\mathcal{O}_{\chi}$ deformation is irrelevant, and we should assume that $\Lambda \gg m_{\gamma} \equiv \frac{e N}{2\pi R} $ to get a well-defined theory without needing to specify a detailed UV completion.  Then the scale relevant to the low-energy physics will be $\sim  m_{\gamma}$, see \eqref{eq:photon_mass}.  When $\Delta_2 < 2$, the $\mathcal{O}_{\chi}$ deformation is relevant, and $\Lambda$ could be larger or smaller than $m_{\gamma}$.   We will focus on the situation where $\Lambda \ll m_{\gamma}$ when $\Delta_2 <2$.  Given this assumption, it will be useful to shift the renormalization scale to $m_{\gamma}$ in \eqref{eq:bosonized_with_gauge_field_mu}  for any value of $\Delta_2>0$.  Using \eqref{eq:scheme_change_finite_g}, this gives the form of the action we will use from here onward:
\begin{align}
  S = \int_{M_2} &\left[ 
   \frac{1}{2 }R^2 \formabs{d\varphi}^2 
   + m_{\gamma}^{\Delta_2}\Lambda^{2-\Delta_2} \cos(2\varphi)  +\frac{1}{2e^2} \formabs{d a}^2  - \frac{i}{2\pi} N d\varphi \wedge  a \right].
  \label{eq:bosonized_with_gauge_field}
\end{align}

As we already mentioned in the introduction, when $m = 0$, and we do not
turn on the $\mathcal{O}_{\chi}$ deformation, the model has $\mathbb{Z}_N$
$1$-form and $0$-form symmetries. These two symmetries have a mixed 't
Hooft anomaly.  One simple way to see this is to analyze the theory on $\mathbb{R}\times S^1$~\cite{Anber:2018jdf}.  Another way is  to write explicit expressions for
the topological operators that generate these
symmetries~\cite{Cherman:2021nox}.  To do this it is helpful to rewrite
the action in first-order form as
\begin{align}
  S_{\textrm{1}^{\textrm{st}} \textrm{ order}} = \int_{M_2} &\left[\frac{1}{2  R^2 }\formabs{b^{(1)}}^2 + i b^{(1)} \wedge d\varphi
  +m_{\gamma}^{\Delta_2}\Lambda^{2-\Delta_2}  \cos(2\varphi) \right.  \nonumber\\
  &\left. +\frac{e^2}{2} \formabs{b^{(0)} }^2  + i b^{(0)} \wedge da - \frac{i}{2\pi} N d\varphi \wedge  a \right].
  \label{eq:bosonized_no_gauge}
\end{align}
The chiral symmetry is associated with the existence of topological line
operators of the form 
\begin{align}
V_k(C) = \exp\left[\frac{2\pi i k}{N} \int_C \left( b^{(1)} + \frac{N}{2\pi} a \right) \right]\,.
\label{eq:chiral_gen}
\end{align}
The $1$-form symmetry is generated by local topological operators $U_n(x)$.  They take form
\begin{align}
U_n(x) = \exp \left[\frac{2\pi i n}{N} \left(b^{(0)}+ \frac{N}{2\pi} \varphi \right)\right].
\label{eq:center_gen}
\end{align}
The key thing to take from these expressions is that $U_n(x)$ is charged under the  $\mathbb{Z}_N$ chiral symmetry, while $V_k(C)$ is charged under the $\mathbb{Z}_N$ $1$-form symmetry.  This means that there is the 't Hooft anomaly between the two $\mathbb{Z}_N$ global symmetries.

The expectation values of $U_1(x)$ take the form
\begin{align}
\langle U_1(x) \rangle = e^{2\pi i k/N} \,,
\end{align}
see e.g.~\cite{Cherman:2021nox} for an extensive discussion.  The choice of $k$ labels the $N$ universes of the model.  Domain walls between universes have infinite tension, and can be thought of as Wilson lines.  The form of Eq.~\eqref{eq:center_gen} implies that chiral symmetry relates different values of $k$.  The different universes have identical vacuum energy densities, which implies that chiral symmetry is spontaneously broken on $\mathbb{R}^2$.    But the fact that domain walls between the chiral vacua can be thought of as Wilson lines implies that the $1$-form symmetry is also spontaneously broken. To summarize,  when $m = 0$ and we do not turn on the $\mathcal{O}_{\chi}$ perturbation, the theory does not confine test charges with $q \in \mathbb{N}$.  

Another instructive perspective~\cite{Komargodski:2020mxz} on deconfinement in the massless
non-deformed Schwinger model is offered by the fact that given a charge-$1$ Wilson loop
$W(C)$ on a contour $C$, we can always insert $V_1(C')$, where $C'$ is a
contour lying e.g. inside $C$, and has an opposite orientation to $C$.
On the one hand, the the operator $V_1(C')$ is topological, so $C'$ can
be shrunk to arbitrarily small size, and $V_1(C' \to 0) \to 1$, so that 
\begin{align}
\langle W(C) \rangle = \langle V_1(C') W(C) \rangle   \,.
\end{align}
 But on the other hand, the operator $V_k(C')$ looks like the world-line of a particle with charge $k$. If we take $C' = \bar{C}$ to be the curve $C$ traversed in the opposite direction, then
 \begin{align}
 \langle V_1(C' = \bar{C}) W(C) \rangle = \left\langle \exp \left[ -\frac{2\pi i}{N} \int_{C} b^{(1)} \right] \right\rangle 
 \end{align}
and the expectation value on the right has a perimeter-law expectation value because $b^{(1)}$ is not electrically charged. This argument leads to us to conclude that $W(C)$ itself must have a perimeter-law expectation value.

If we turn on the $\mathcal{O}_{\chi}$ deformation, so that $\Lambda
\neq 0$, the exact chiral symmetry is reduced to $\mathbb{Z}_2$.  (The
approxiate low-energy chiral symmetry can be larger.) There is now
only one topological line operator, $V_{N/2}(C)$.    Correspondingly,
the 't Hooft anomaly is between the surviving $\mathbb{Z}_2$ chiral
symmetry and the $\mathbb{Z}_2$ subgroup of center symmetry  generated
by
\begin{align}
U_{N/2}(x) = \exp \left[i\pi\, b^{(0)}+ \frac{iN}{2} \varphi \right]  \,.
\end{align}
This means that test charges with $q = N/2$ should be deconfined.   The
fate of confinement for test charges in other representations depends on
whether $\mathcal{O}_{\chi}$ is relevant, and will be
discussed below. 

\section{Dynamics on $\mathbb{R}^2$}
In the following sections we examine the dynamics of the charge-$N$ Schwinger model with
four-fermion deformations.  We first discuss the physics on $\mathbb{R}^2$, where
our analysis will be under analytic control so long as the coefficient of the
$\mathbb{Z}_2$-invariant chiral symmetry-breaking four-fermion deformation is
small enough. 

First, suppose $m=0$, and turn off the $\mathcal{O}_{\chi}$ deformation.  Let us view $\mathbb{R}^2$ as the infinite-volume limit of some closed manifold $M_2$, such as a torus, and drop the boundary term in Eq. \eqref{eq:bosonized_with_gauge_field}.  The gauge field $a$ enters the action as $f = da$, so instead of integrating over $a$ we can integrate over $f$ as long as we ensure that the fluxes of $f$ on $M_2$ are properly quantized.    We integrate out $b^{(0)},b^{(1)}$ in Eq. \eqref{eq:bosonized_with_gauge_field} to get an action in terms of $\varphi$ and $a$, and then note that the partition function can be written as
\begin{align}
  Z &= \int \mathcal{D}\varphi\, \mathcal{D}f \, \sum_{\nu \in \mathbb{Z}}\delta\left(\nu - \frac{1}{2\pi} \int_{M_2} f\right) e^{-S_{\rm Schwinger}} \nonumber\\
  &= \int \mathcal{D}\varphi\, \mathcal{D}f\, \sum_{k \in \mathbb{Z}} \exp\left(i k \int_{M_2} f\right) e^{-S_{\rm Schwinger}} \nonumber \\
  &= \sum_{k \in \mathbb{Z}} \mathcal{D}\varphi\, e^{-\int_{M_2} \mathcal{L}_k} \,,
\label{eq:Zvarphi}
\end{align}
where 
\begin{align}
  \mathcal{L}_k &= \half R^2  \formabs{d \varphi}^2 + 
  \frac{1}{2} \left(\frac{e N}{2\pi}\right)^2 \formabs{\varphi - \frac{2\pi k}{N}}^2 = \frac{1}{2} \left[ \formabs{d \tilde \varphi}^2
  +m_{\gamma}^2 
 \formabs{\tilde\varphi - \frac{2\pi k\, R }{N} }^2 \right]\,.
\end{align}
In the second equality above we switched to the canonically normalized field $\tilde
\varphi$, $\varphi =\tilde\varphi/R$, and defined
\begin{align}
m_{\gamma} =  \frac{e N}{2\pi R} \,,
\label{eq:photon_mass}
\end{align}
which reduces to $m_{\gamma} = e N/\sqrt{\pi}$ when $g = 0$.  The sum over $k$ in Eq.~\eqref{eq:Zvarphi} ensures that the path integral is invariant under $\varphi \to \varphi +2\pi$.  Finally, we take the limit $M_2 \to \mathbb{R}^2$ in Eq.~\eqref{eq:Zvarphi}.

The periodicity of $\tilde{\varphi}$ is
$2\pi R$.  In the limit $g \to \infty$ with $e N$ fixed, the mass gap vanishes,
and the long distance local physics is that of an $\mathbb{R}$-valued massless
scalar.  We can think of $k$ as a universe label (mod $N$), and the universes
are all degenerate.  So as long as the $\mathcal{O}_{\chi}$ operator is turned off, the $\mathbb{Z}_N$ $1$-form symmetry is
spontaneously broken for any $g$, as is the $\mathbb{Z}_N$ $0$-form chiral symmetry.

Now consider turning on the deformation by $O_\chi$, while keeping $m=0$.  Consulting
Eq.~\eqref{eq:bosonized_no_gauge} we get
\begin{align}
  \mathcal{L}_{k} =\half R^2 \formabs{d \varphi}^2 + 
  \frac{1}{2} \left(\frac{e N}{2\pi}\right)^2 \formabs{ \varphi - \frac{2\pi k}{N} }^2 - m_{\gamma}^{\Delta_2} \Lambda^{2-\Delta_2} \cos(2\varphi)
\end{align}
When $R^2 = 1/4\pi$ (the $g=0$ point), the dimensionless factor in the coefficient
of $\cos(2\varphi)$ is $ \left(\frac{eN}{\Lambda} \right)^2$, so as we take
$\Lambda \gg e N$ the $\cos(2\varphi)$ term becomes less and less important.  This
is consistent with the expectation that $\cos(2\varphi)$ is irrelevant at $g=0$.
On the other hand, if $R^2 > 1/2\pi$ (corresponding to $g > \pi/2$), the
$\cos(2\varphi)$ term becomes large when $\Lambda \gg e N$, which is consistent
with the expectation that $\cos(2\varphi)$ becomes relevant when $g > \pi/2$.  

When $\Delta_2 > 2$ the $\mathcal{O}_{\chi}$ perturbation is not all that interesting.  The scale $\Lambda$ is a short-distance scale, past which the theory needs a UV completion. We we get an emergent $\mathbb{Z}_N$ chiral symmetry at long distances as we take the UV scale $\Lambda$ to infinity.  The considerations in Sec.~\ref{sec:elementary_confinement} imply that there is a non-vanishing string tension for $m_{\gamma}/\Lambda \neq 0$, but it goes to zero as $\Lambda$ becomes large.  In this sense the theory is
no more (and no less) confining than compact $U(1)$ QED in $2+1$ dimensions on
a Euclidean lattice with the Wilson gauge action.



For us the more interesting case is $R^2 > 1/(2\pi)$, where $\mathcal{O}_{\chi}$ is relevant, and $\Lambda$ should be interpreted as an infrared scale, which we assume is small compared to $m_{\gamma}$.\footnote{If $R^2 = 1/(2\pi)$, then $\mathcal{O}_{\chi}$ is marginal if the gauge interaction is turned off and $m = 0$.  This means that $\mathcal{O}_{\chi}$ enters the action with a dimensionless coefficient which we can call $\lambda$.  It may be interesting to understand whether and how this parameter runs with the renormalization scale.}  If $N$ is even, then the chiral symmetry is $\mathbb{Z}_2$ at long distances.  
We would like to determine the realization of
this symmetry, as well as the realization of the $\mathbb{Z}_N$ 1-form symmetry.
To do this it is important to understand when our Lagrangian is weakly coupled.
The squared mass of the particles created by $\varphi$ in the $k$th universe
(that is, the coefficient of $\frac{1}{2} \tilde{\varphi}^2$)  is
\begin{align}
m_{\rm eff}^2 =m_{\gamma}^2  \left[1  
+  \left(\frac{\Lambda}{m_{\gamma}}\right)^{2-\Delta_2} 
\frac{4}{R^2}  
 \cos \left(\frac{4 \pi  k}{n}\right) + \cdots \right]\,.
\end{align} 
while the coefficient of e.g. 
the quartic interaction $\frac{\lambda}{4!}
\tilde{\varphi}^4$ is
\begin{align}
\lambda = m_{\gamma}^2  \left[ -  \frac{16}{R^4} \left(\frac{\Lambda}{m_{\gamma}}\right)^{2-\Delta_2} \cos\left(\frac{4\pi k}{N}\right) + \cdots \right]\,.
\end{align}
where the $\cdots$ represents terms that are higher-order in the small parameter $\left(\frac{\Lambda}{m_{\gamma}}\right)^{2-\Delta_2} $.
The theory is weakly-coupled if e.g. the parameter $|\lambda/m_{\rm eff}^2| \ll 1$, and 
\begin{align}
\left|\frac{\lambda}{m_{\rm eff}^2}\right| &= 
\frac{16 \left(\frac{\Lambda}{m_{\gamma}}\right)^{2-\Delta_2}  \cos \left(\frac{4 \pi  k}{n}\right)}{R^4 \bigg[1+  \frac{4}{R^2} \left(\frac{\Lambda}{m_{\gamma}}\right)^{2-\Delta_2} \!\! \cos \left(\frac{4 \pi  k}{n}\right) \bigg]} \simeq \frac{16}{R^4} \left(\frac{\Lambda}{m_{\gamma}}\right)^{2-\Delta_2}  \cos \left(\frac{4 \pi  k}{n}\right)\,.
\end{align}
This illustrates the fact that the massless four-Fermi-deformed Schwinger model is weakly coupled when $\left(\frac{\Lambda}{m_{\gamma}}\right)^{2-\Delta_2}  \ll 1$, just as the conventional massive Schwinger model is weakly coupled when $m_{\psi}/e \ll 1$.  

When the model is weakly-coupled we can read off the vacuum structure and the confining string tensions by considering the vacuum energy densities of the
universes in the Lagrangian.  Indeed, this was already discussed in Sec.~\ref{sec:confinement_elementary}, where we had already tacitly assumed that there is a duality frame where the theory is weakly coupled.

\section{Dynamics on $\mathbb{R} \times S^1$} 
\label{sec:circle_physics}

We now discuss the calculation of the string tension both in the mass-perturbed
theory as well as the massless theory with four-fermion perturbations on $\R
\times S^1 $ with small $S^1$, as well as on $T^2$.  The results below match the
expectations established in our analysis of the physics on $\mathbb{R}^2$.   The
benefit of working out the physics on a small circle is that it allows us to
establish a semiclassical picture of the confinement mechanism.  
In particular, we will show below that confinement is induced by the proliferation of fractional instantons with
topological charge $Q = \pm 1/N$ and action $S_1 =1/(4R^2m_\gamma L)$.  These instantons
all carry fermion zero modes when the theory has a $\mathbb{Z}_N$ chiral
symmetry.  So when the complete chiral symmetry is present, the model does not confine
$q = \pm 1$ test charges.  When chiral symmetry is explicitly broken,
either by a mass term or by the $\mathcal{O}_{\chi}$ deformation, some
or all of the fermion zero modes get lifted, and then confinement 
sets in.  For earlier discussions of
instantons in various versions of the Schwinger model see e.g. Refs.~\cite{Jayewardena:1988td,Sachs:1991en,Smilga:1993sn,Shifman:1994ce,Hetrick:1995wq,Smilga:1996dn,Rodriguez:1996zj,Radozycki:2008zt,Misumi:2019dwq}.

Our analysis below has several unusual features.  The first one is
already hinted at above: we will see that the massless
Schwinger model on a cylinder has instanton solutions with topological charge
$|Q| = \ell/N$ and actions $S = \ell^2 S_1$, which carry $2\ell$ fermion zero
modes when the model has $\mathbb{Z}_N$ symmetry.   Remarkably, there are also
exact solutions of the equations of motion with \emph{zero} topological charge
that carry \emph{exact} fermion zero modes.  These $Q = 0$ solutions are obtained
simply by adding together solutions with non-zero topological charge.  The fact
that one can obtain exact solutions this way comes from the fact that in the
bosonized duality frame, the equations of motion satisfied by the instantons are
\emph{linear} in the fields.  The fact that instantons with $|Q| \ge 0$ carry
fermion zero modes implies that they do not contribute at all to the partition
function of the massless Schwinger model with $\mathbb{Z}_N$ symmetry.  Of
course, as soon as we consider correlation functions of operators that involve
the fermions, or add chiral-symmetry breaking deformations to the action, the
story changes and the instantons do contribute.
 
\subsection{Fractional instantons}
We begin our analysis on $\mathbb{R} \times S^1$ with the massless charge-$N$
Schwinger model with $\mathbb{Z}_N$ chiral symmetry.  We denote the coordinate
of $S^1$ by $x$, and denote the coordinate of $\mathbb{R}$ by $\tau$.  When the
circumference of $S^1$, $L$, is small enough compared to $1/e$, we can use an
EFT on $\mathbb{R}$ --- that is, quantum mechanics --- to describe the
long-distance dynamics.  To describe this EFT we need to understand the
potential for the gauge field holonomy $\Omega_x$ on $S^1$. If we choose periodic boundary
conditions for the charge $N$ fermion,\footnote{Physically, the boundary
condition on the fundamental fermion does not matter in the Schwinger model,
because it is part of gauge redundancy. Any boundary condition  $\psi(x_1, x_2+
L_2) = e^{i \alpha } \psi(x_1,x_2)$ will yield the same physical result.  } and
take Coulomb gauge such that $\Omega_x = e^{i \int dx a_x} \equiv e^{i h(\t)}$,
where $h \sim h+2\pi$, the 1-loop holonomy potential is given by 
\begin{align}
V(h)= \frac{2}{\pi L^2} \sum_{n=1}^{\infty} \frac{1}{n^2} \cos(N n h) 
= \min_n \frac{N^2}{2\pi L^2}\left(h - \frac{2\pi}{N} (n+\tfrac{1}{2})\right)^2 - \frac{\pi}{6L^2}\, . 
\label{hol-pot}
\end{align}
The fundamental domain $h \in [0,2\pi)$ contains $N$ harmonic minima located at
\begin{align}
h_n = \frac{2 \pi}{N}\left(n+  \textstyle \frac{1}{2} \right),  \qquad n=0, 1, \ldots N-1\,.
\label{min-pot}
\end{align} 
If we take Eq.~\eqref{hol-pot} as a quantum effective potential in the 1d EFT
description of the system valid for small $L$, then naively this EFT will
support $N$ distinct instanton solutions interpolating between consecutive minima: 
\begin{align}
{\cal F}_{n+1}:  \qquad | n \rangle_B  \rightarrow | {n+1} \rangle_B\,,  \qquad n=0, 1, \ldots N-1,  \;\; N \equiv 0\,.
\label{min-FI}
\end{align}

To understand these instantons, we first recall that the topological charge of $U(1)$ gauge theory is defined as
\begin{align}
  Q = \frac{1}{2\pi}\int_{M_2} f
= \frac{1}{4\pi}\int d^2x\, \epsilon^{\mu\nu} f_{\mu \nu} \,.
\end{align}
In pure $U(1)$ gauge theory $Q$ is an integer, and on e.g. $T^2 = S^1_{\beta}
\times S^1_{L}$ we can write instanton solutions as e.g. $a_{\tau} =0, \; a_{x} = \frac{2\pi Q
\tau}{L\beta} $\,.  The action of a charge $Q$ instanton is 
\begin{align}
  S = \frac{1}{4 e^2} 2 L \beta  \left(\frac{2\pi^2 Q}{L\beta}\right)^2 = \frac{2\pi Q^2}{e^2 L \beta}\,.
\end{align}
These $Q \in \mathbb{Z}$ instantons are of course also present in $U(1)$ gauge
theory with matter. However, note that in a $Q \in \mathbb{Z}$ instanton,
$\int_{S^1_L} a$ evolves from $0$ to $2\pi Q$ as $\tau$ goes from $0$ to
$\beta$.   This means that $Q \in \mathbb{Z}$ instantons cannot represent
tunneling events between the nearest-neighbor vacua of \eqref{hol-pot}.

What we need are instantons with fractional topological charge, with $Q = \ell/N$, with $\ell \in
\mathbb{Z}$. When $\ell = 1$, these fractional instantons are
precisely the tunneling solutions of Eq.~\eqref{min-pot}.  In the absence of
fermion zero modes, these fractional instantons can only directly contribute to
the partition function if one gauges the $\mathbb{Z}_N$ $1$-form symmetry.  But
when the $\mathbb{Z}_N$ $1$-form symmetry is \emph{global}, the topological
charge of admissible field configurations that contribute to the partition
function must be an integer, and the partition function only receives
contributions from `composite' instantons with $Q \in \mathbb{Z}$ built from
the field configurations with the fractional charges $Q =  \ell/N$.  For a
general discussion of how fractional-charge instantons contribute to
gauge-theory path integrals see Ref.~\cite{Unsal:2020yeh}.

However, the effective potential in Eq.~\eqref{hol-pot}  has cusps, so we have
to be careful when using it to study tunneling solutions.  Equation \eqref{hol-pot}
should be understood as the 1-loop vacuum free energy density in a constant
holonomy background, extracted by taking a large-volume limit. This is to be
distinguished from the 1-loop effective potential in the small-$L$ quantum
mechanical theory describing the \emph{dynamics} of the holonomy. Indeed, the
fact that the energy density has cusps indicates that the effective 1d EFT fails
to capture important physics---in this case, the existence of fermion zero
modes.  This follows from the standard ABJ anomaly which is responsible for the
chiral symmetry being $\mathbb{Z}_N$ rather than $U(1)$.  This anomaly implies
that field configurations with topological charge $Q$ have $2N$ fermion zero
modes.

To better understand the tunneling solutions and their zero modes, we can
analyze the small-$L$ limit using bosonization, following
Ref.~\cite{Misumi:2019dwq}. We work on $T^2 = S^1_\beta \times S^1_L$, with $L
\ll 1/e \ll \beta$, and take the coordinates on the torus to be $(\tau, x) \sim
(\tau + \beta, x+L)$. Our goal is to understand the effective action for
$\Omega_x = e^{i \int dx\, a_x}$, which is derived in detail in
Appendix~\ref{sec:appendix}. Just as above, we take Coulomb gauge such that
$\Omega_x \equiv e^{i h(\t)}$, where $h \sim h+2\pi$.

The interesting contribution to the physics comes from the sum over the scalar winding number $\oint_{S^1_\beta} d\varphi \in 2\pi \mathbb{Z}$. The small-$L$ path integral capturing the dynamics of the holonomy
becomes
\begin{align} \label{eq:smallcircleZ}
Z \sim  \int \mathcal D h \,\exp\left[-\int d\tau \frac{1}{2e^2L}\left(\frac{dh}{d\tau}\right)^2\right] \sum_{n\in\ZZ} \exp\left[-  \int d\tau\, \frac{m_\gamma^2}{2e^2 L}\left(h-\frac{2\pi n}{N }\right)^2 \right]\,.
\end{align}
There are two things to notice about the above expression. First, if we take
$h=$ constant and take the large-volume limit $\beta \to \infty$, the above
effective potential reduces to the vacuum free energy density in
Eq.~\eqref{hol-pot}, up to an $h$-independent shift. Second, and more
importantly, the effective potential for the holonomy consists of $N$ distinct
branches. Each of the $N$ minima, located at $h = 2\pi n/N$, lie in distinct
branches of the effective potential. 

An immediate consequence of the discussion above is that tunneling
configurations $h(\tau)$ between minima of the effective potential simply do not
exist.  In the fermionic duality frame, this statement maps to the fact that
instantons carry robust fermion zero modes, and so they cannot contribute to the
partition function.   To allow tunneling events between minima, we must insert
an operator charged under chiral symmetry.  This is the bosonic analog of
`soaking up' the fermion zero modes. The necessary operator is simply $e^{i \ell
\varphi}$, which carries charge $\ell$ under the $\mathbb{Z}_N$ chiral symmetry.
The fermionic-variable image of $e^{i \ell \varphi}$ can be thought of as either a point-split version of
$\left(\bar{\psi}_L \psi_R\right)^\ell$, or its local operator analogue with derivatives. 

\begin{figure}[t]
  \vspace{-1.cm}
  \begin{center}
  \includegraphics[width = 0.75\textwidth]{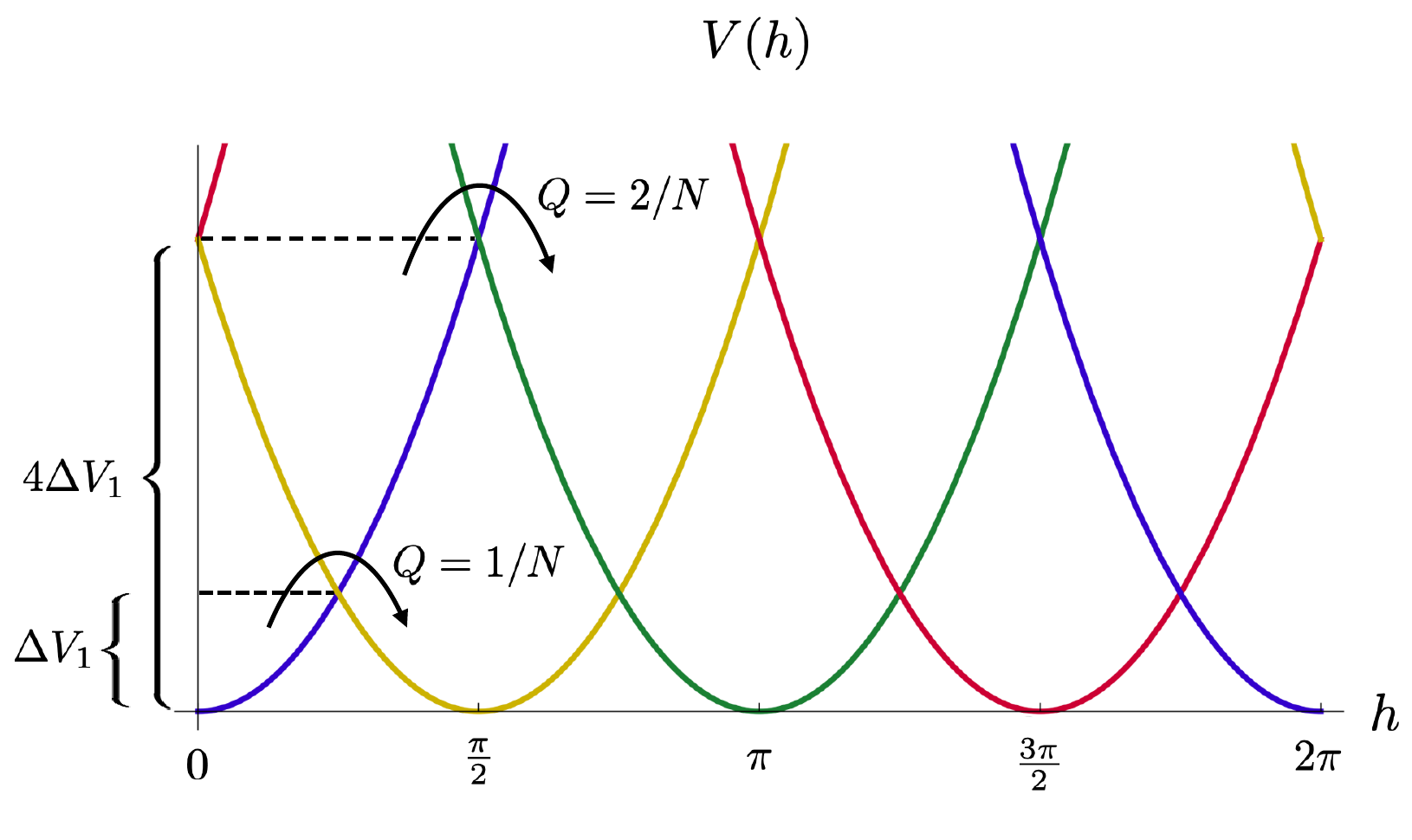}
  \caption{In the bosonized description, tunneling events connecting different
      branches of the holonomy effective potential are only possible in the
      presence of insertions carrying chiral charge. In the presence of the
      insertion $e^{i\ell\varphi(\tau_0,x_0)}$, a tunneling
      configuration in the $n$th branch starts at $h=2\pi n/N$ and travels up
      the potential until it meets the $(n+\ell)$th branch at time $\tau_0$.
      At this point, the insertion of the chirally-charged operator rearranges
      the branches and the tunneling configuration descends down to $2\pi
      (n+\ell)/NL$. In the above figure, we took $N=4$ and the cases $\ell=1,2$
      are indicated. The height that the instanton has to traverse scales as
      $\ell^2$, as does the instanton action \eqref{eq:instantonaction}. }
  \label{fig:tunneling}
  \end{center}
\end{figure}

Repeating the derivation of the holonomy effective potential in the presence of the insertion $e^{i\ell
\varphi(\tau_0,x_0)}$, one finds (see Appendix~\ref{sec:GPY_derivation})
\begin{equation} \label{eq:GPY_insertion}
\tilde Z \sim \int \mathcal D h \,\exp\left[-\int d\tau \frac{1}{2e^2L}\left(\frac{dh}{d\tau}\right)^2\right] \sum_{n\in\ZZ} \exp\left[-  \int d\tau \frac{m_\gamma^2}{2e^2L}\left(h-\frac{2\pi (n+\ell \, \Theta(\tau-\tau_0))}{N}\right)^2 \right]\,.
\end{equation}
The equation of motion for $h$ in the presence of the insertion becomes
\begin{equation}
\frac{d^2h}{d\tau^2} = m_\gamma^2 \left( h - \frac{2\pi }{N}\left(n+\ell\,\Theta(\tau-\tau_0)\right)\right),
\label{eq:instanton_eq}
\end{equation}
in the $n$th branch. This equation has finite-action solutions interpolating
between $h = 2\pi n/N$ and $2\pi (n+\ell)/N$ which are illustrated in
Fig.~\ref{fig:tunneling} and take the form
\begin{equation}
h_{n,\ell}(\tau;\tau_0) = \frac{2\pi n}{N} + 
\begin{cases}
\frac{2\pi \ell}{N}\frac{\cosh(m_\gamma  (\tau_0-\beta))}{\sinh(m_\gamma \beta)}\sinh(m_\gamma  \tau) &\quad \tau \leq \tau_0 \\
\frac{2\pi \ell}{N}\left[1+ \frac{\cosh(m_\gamma \tau_0)}{\sinh(m_\gamma  \beta)}\sinh(m_\gamma  (\tau-\beta))\right] &\quad \tau > \tau_0
\end{cases}\,.
\end{equation}
The topological charge and action of such an instanton in the $\beta \to \infty$
limit are
\begin{equation}\label{eq:instantonaction}
Q = \frac{\ell}{N}, \quad S_{\rm \ell} = \frac{ \ell^2}{4R^2 m_\gamma L}\,. 
\end{equation}

The $h$ equation of motion \eqref{eq:instanton_eq} with a finite number of
insertions of $e^{i\varphi}$ is linear, so sums of solutions are also solutions.
For example,
\begin{align}
h_{n;Q = 0}  =  h_{n,\ell}(\tau;\tau_0) + h_{n+1,-\ell}(\tau + \tau';\tau_0 +\tau') + \frac{2\pi \ell}{N}
\label{eq:zero_chare_instanton}
\end{align}
is an exact solution of the $h$ equation of motion with an insertion of
$e^{i\ell \varphi(\tau_0,x_0)}$ and $e^{-i\ell \varphi(\tau_0+\tau',x_0)}$ with
$Q = 0$ and an action $S = \frac{\ell^2}{2R^2 m_\gamma L}$ when $\beta$ is
large.  This solution has two unusual features.  First, it is already quite peculiar to
have an exact real solution of the equations of motion with $Q = 0$.  Note that Eq.~\eqref{eq:zero_chare_instanton} is \emph{not} a saddle point at infinity (see e.g.~\cite{Behtash:2018voa}), since $\tau'$ is finite.  The other unusual feature is that this $Q = 0$
solution has exact fermion zero modes, which are localized at
$\tau = \tau_0$ and $\tau_0+\tau'$.  This means that despite having vanishing topological charge it does not contribute to the partition
function of the massless charge-$N$ Schwinger model with $\mathbb{Z}_N$ chiral
symmetry.

\subsubsection{Fermion zero modes and confinement}

Let us pass back to the fermionic description of the Schwinger model.
The fermion zero modes of the fractional instantons discussed above play
a crucial role in determing the dynamics of the charge-$N$ Schwinger
model with a $\mathbb{Z}_N$ chiral symmetry. The holonomy effective
potential has $N$ degenerate minima, so that the theory has an $N$-fold
degenerate vacuum on a cylinder at the classical level. Normally one
might expect this degeneracy to be lifted due to tunneling events
connecting the classical vacua. However, our analysis above shows that
tunneling events between these minima are forbidden, and the $N$-fold
degeneracy of vacua persists when quantum effects are taken into
account.  This is due to the mixed 't Hooft anomaly between
center-symmetry and chiral symmetry, and in practice, the absence of
tunneling is enforced  via the fermionic zero mode structure of the
fractional instanton events.

\begin{figure}[t]
  \vspace{-0.5cm}
  \begin{center}
  \includegraphics[width = 0.9\textwidth]{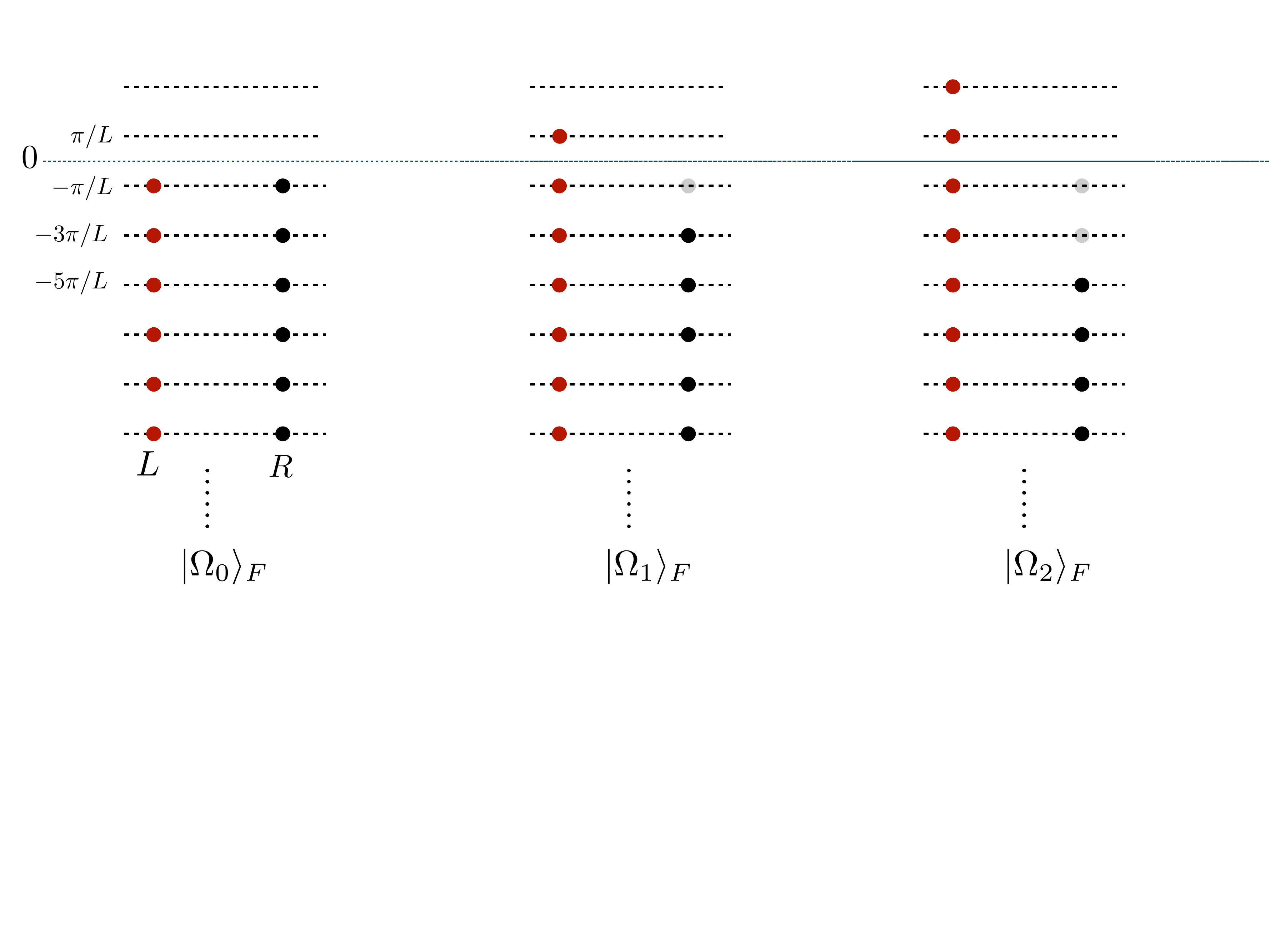}
  \vspace{-4.0cm}
  \caption{The fermionic vacua  $|\Omega_n \rangle_F$ of the theory corresponding to  harmonic 
   bosonic  ground state $|n \rangle_B $ where  $h_n=  \frac{2 \pi}{N}\left(n+  \textstyle \frac{1}{2} \right)$, and 
   $\Delta h= \frac{2 \pi}{N}$. The change in the bosonic background sources is $\Delta Q_5=2 $.
      }
  \label{fig:Fermistates}
  \end{center}
\end{figure}

In the fermionic description, the harmonic perturbative vacua  are not only described via the bosonic states $
| n \rangle_B$, but also via degenerate fermionic states, which change between
adjacent harmonic minima. The state $|n\rangle$ associated with the $n$-th vaccum is a vector
in the tensor product of
states in bosonic and  fermionic Hilbert spaces: 
\begin{align}
| n \rangle  \equiv  | n \rangle_B   \otimes | \Omega_n \rangle_F
\label{har-BF}
\end{align}
We can take  $| \Omega_0 \rangle_F$  to be all L- and R-handed negative energy
levels filled up as shown in Fig.~\ref{fig:Fermistates}. $ | \Omega_1 \rangle_F $
differs from it by $\Delta Q_5= 2$ where  all L states are shifted up by one
unit and all R states are shifted down by one unit, i.e. a R-handed state
is removed and a L-handed state is created.   This is due to the  ABJ anomaly
and the consequent zero mode structure of fractional instantons.  The usual
instantons in the theory have  $2N$ fermion zero modes.  However, the fractional
instantons interpolating between consecutive vacua  \eqref{min-FI} have
topological charge $Q=1/N$, leading to  $\Delta Q_5= 2$.\footnote{A Hilbert
hotel analogy is useful. One can think of both $L$ and $R$ as  Hilbert hotels.
Both are infinite, and  can always make a vacancy  by moving all visitors to the
next room. Or when a visitor is pushed out, the hotel still remains full. }
As a result, the degenerate fermionic state   $| \Omega_n \rangle_F$ can be written as: 
\begin{align}
 | \Omega_n \rangle_F = \prod_{k=-n}^{\infty}  |k \rangle_L \otimes  \prod_{k=-\infty}^{-n-1}  |k \rangle_R \end{align} 
as shown in Fig.~\ref{fig:Fermistates}.

 The amplitudes of the fractional instantons are of the form 
 \begin{align}
 {\cal F}_j \sim  e^{-S_1} e^{i \theta/N}  \overline \psi_L  \psi_R, \; \;\;  j=1, \ldots, N \,.
 \end{align}
 The fermion fields in the prefactor of the instanton amplitude represent the
 fermionic zero modes.  In the bosonized description, these zero modes are
 represented by the need to insert the operator $e^{i\varphi}$ at a spacetime
 point $x$ to enable a $Q = 1/N$ tunneling event centered at $x$.  Thanks to the
 fermionic zero modes, the $Q = 1/N$ instantons do not lift the degeneracy
 between the $N$-vacua, since  the transition amplitude 
 \begin{align}
 \langle n+1 | e^{-\beta H} | n \rangle =0
 \end{align}
 remains zero.  Hence, as stated earlier, charge-$N$ QED with massless
 fermions and no four-fermion deformations has $N$ exactly degenerate
 bosonic vacua on  $\R^1 \times S^1$.   This can be viewed as a
 consequence of the mixed 't Hooft anomaly between the $\mathbb{Z}_N$
 chiral symmetry and the $\mathbb{Z}_N$ $1$-form symmetry, see e.g.
 Appendix D of~\cite{Gaiotto:2017yup} for a simple example of this sort
 of phenomenon.

\subsection{Mass perturbation}

It has long been known that adding a mass term $m_{\psi} \bar{\psi}\psi$
induces confinement.  Of course, it also lifts  the degeneracies between the $N$ ground states of the
$m_{\psi} = 0$ theory.  Indeed, when $S^1$ is small, and $0< m \ll e$,  the nearest-neighbor matrix elements become
non-zero, because the mass term can be used to ``soak up'' the fermion zero modes,
\begin{align}
\langle n \pm 1  | e^{-\beta (H + \Delta H_{\rm mass}) } | n \rangle 
= \mu\, m L \beta K_1 \,e^{-S_1  \pm  i \frac{\theta}{N}} 
\label{eq:mass_tunneling} 
\end{align}
where $K_1 =\tfrac{1}{2} \,$, and we have set $g=0$ for simplicity.  More generally the powers of the parameters in the prefactor in the tunneling amplitude depend on the scaling dimension of $\bar{\psi}\psi$, which depends on $g$.
Indeed, in the bosonized description of the model, we can formally expand the 
\begin{align}
  \exp{\left[\int
d^{2}x\,\mu \,m  \left(e^{i\varphi(x)}+e^{-i\varphi(x)}\right)\right]}
\end{align}
 term in the exponentiated Euclidean action in powers of $m$, and then note that
this produces a sum of powers of $e^{\pm i \varphi(x)}$ summed over the insertion points $x$.  These insertions induce tunneling events from $\varphi = 2\pi n/N$ to $\varphi = 2\pi(n \pm 1)/N$.  Then
the analysis in the preceding section implies that the $\ell = \pm 1$ events
contribute to the path integral with the weights given in
\eqref{eq:mass_tunneling}.

To see the effect of the breaking of degeneracy of the vacua on confinement, let us consider
the partition function  $Z(\beta)$ of the system in Born-Oppenheimer
approximation. To calculate it, we need to sum over all  periodic paths,
$a(\beta)= a(0)$.   These paths are  described by maps $S^1 \rightarrow S^1$,
and  
are classified by the winding number $W  \in \pi_1(S^1) = \Z$, which is the
integer valued topological charge. On the other hand, the physical system
possess topological configurations with fractional topological charge $Q= \pm
1/N$. They are seeded by the $e^{\pm i \varphi(x)}$ operators that come from the
expansion of the mass term. If $n$ $(\bar n)$  denotes the number of fractional
instantons and anti-instantons, the configurations that contribute to $Z(\beta)$
must satisfy 
\begin{align}
  n -\bar n = W N, \, W \in \Z .
  \label{eq:charge_condition}
\end{align}
This condition is enforced
by the constraint on the winding number.  It is also enforced by the fact that
when we expand the mass term in powers of $m$ to produce a sum over insertions
of $e^{\pm i \varphi}$, the path integral measure is invariant under
$\mathbb{Z}_N$ chiral symmetry.  Equation \eqref{eq:charge_condition} allows configurations with
integer topological charge  but fractional action  to contribute, e.g.,  $W=
\frac{1}{N}- \frac{1}{N}=0, S=2S_1$ where $S_1 \sim 1/N$.  As a result, the
partition function can be expressed as:
  \begin{align} 
 Z(\beta, \theta) 
 &= N  
 \sum_{W \in \Z}  \sum_{n=0}^{\infty}   \sum_{\bar n =0}^{\infty}   \frac{1}{n!} \frac{1}{\bar n !}  \left( \beta L \mu m K_1  e^{-S_1 + \im \theta/N}  \right)^{n}  \left(  \beta L \mu m K_1  e^{-S_1 - \im \theta/N}  \right)^{\bar n}    \delta_{n-\bar n  - W N, 0}  \nonumber \\
 &=  \sum_{k =0}^{N-1}   \sum_{n=0}^{\infty}   \sum_{\bar n =0}^{\infty}   \frac{1}{n!} \frac{1}{\bar n !}  \left(  \beta L \mu m K_1  e^{- S_1 + \im  \frac{ \theta+ 
2 \pi k} {N}}  \right)^{n }     \left(  \beta L \mu m K_1  e^{- S_1 -  \im \frac{\theta+ 2 \pi k} {N}}  \right)^{\bar n }  \nonumber \\
 &=   \sum_{k =0}^{N-1}   e^{  2 \beta  K_1  e^{- S_1} \cos  \frac{ \theta+ 2 \pi k} {N} }   =   \sum_{k =0}^{N-1}   e^{ - \beta E_k(\theta)}\,. \label{PF}
\end{align}
Here we have assumed the dilute-instanton-gas limit, where $S_1 \gg 1$.  To pass to the second line we converted  the sum over winding number  $W \in \Z$   into
a sum over  lattice momenta  
$k$ that  form eigenstates of $\Z_{N}$ translation symmetry by using the
identity $ N \sum_{W \in \Z} \delta_{n-\bar n  - WN}  =   \sum_{k=0}^{N-1}  
  e^{ \im 2 \pi   k (n-\bar n)/N}  $.  This can be used to do the sum over $n$  and
  $(\bar n)$, and yields the energy eigenvalues of the
  $N$ branches of the vacuum energy in the small $m_{\psi}$ limit: 
\begin{align}
   E_k(\theta) = -\mu\, mL\,  e^{- S_1} \cos \left( \frac{ \theta+ 2 \pi k} {N} \right)\,.
  \end{align} 
    The corresponding eigenstates of the Hamiltonian  are given by 
      \begin{align}
|\theta, k \rangle =  \sum_{n \in \Z} e^{ i \frac{ \theta+ 2 \pi k}{N} n} | n \rangle \,.
\end{align}
 The ground state energy in a given range of theta is found by minimizing over
 the branches, and   is given by $E_{\rm gr}(\theta) = {\rm Min}_k
 E_k(\theta)  $. In the $-\pi/N < \theta < \pi/N$ range, the ground state is the
 $k=0$ branch. 
  
 We can now compute the string tension. Computing the two-point function of
 charge-$k$  Polyakov loops at a large separation $\tau$ in the vacuum for
 $-\pi < \theta < \pi$ which correspond to $|\theta, k'=0\rangle$,    we obtain 
 \begin{align}
&\langle\theta,0| P^{-k}(\tau) P^k(0)|\theta, 0\rangle   \sim \exp\left[-\tau (E_k(\theta)-E_0(\theta))\right].
\label{cor}
\end{align}
In the semi-classical regime, we obtain
 \begin{align}
T_k(\theta) = -  \mu\,m\, e^{-\pi/m_\gamma L}\left\{\cos\left({\theta+2\pi k \over N}\right)-\cos\left({\theta\over N}\right)\right\} .                                                                                                                                                                                                                                                                                                                                                                                                                                                                                                                                 
\end{align}
Note that in the semi-classical domain, as on $\R^2$, the string tension
vanishes   for massless fermions. The benefit of working on $\R\times S^1$ is
that we can see the mechanism of confinement in the massive charge-$N$ Schwinger
model:  it is induced by fractional instantons with $Q = 1/N$, provided that
their fermion zero modes are lifted.

\subsection{$\mathcal{O}_{\chi}$ perturbation}

We now discuss the other perturbation that can lift the vacuum degeneracies of
the charge-$N$ Schwinger model:  the perturbation by the operator
$\mathcal{O}_{\chi}$.  When $N$ is even, this operator breaks the $\Z_N$ chiral
symmetry to $\Z_2$.  In the fermionic description, this is a four-fermion
operator.  This operator can be relevant or irrelevant depending on the
coefficient of the only chirally-invariant perturbation $\mathcal{O}_{\rm jj}$ of the
model.  Of course, the deformation by $\mathcal{O}_{\chi}$ is most interesting
when it is relevant.

When  $\mathcal{O}_{\chi}$ is added to the action and $N$ is
even, then
\begin{align}
\langle n \pm 1 | e^{-\beta (H + \Delta H_{\chi})} | n \rangle &=0\,,
\label{eq:NN_chi_even}\\
\langle n \pm 2  | e^{-\beta (H + \Delta H_{\chi}) } | n \rangle 
&= \beta L K_2 \Lambda^{2-\Delta_{\chi}} \mu^{\Delta_{\chi}}\, e^{-S_{2} \pm  2 i \frac{\theta}{N}}  \,.
\label{eq:NNN_chi_even}
\end{align}
where $K_2$ is a dimensionless constant. To see this we can formally expand the $\exp{\left[\int\! d^{2}x
\,\Lambda^{2-\Delta_{\chi}} \mathcal{O}_{\chi} \right] }$ term in the exponentiated
Euclidean action in powers of $\Lambda$, and then note that this produces a sum
over insertions of $e^{\pm i \ell \varphi(x)}$ with $\ell \in 2\mathbb{Z}$, which
induce tunneling events localized at the points $x$.  The absence of insertions
with $\ell = \pm 1 \textrm{ mod } N$ implies \eqref{eq:NN_chi_even}, while
the presence of insertions with $\ell = \pm 2$ implies \eqref{eq:NNN_chi_even}.

If $N$ is odd, then after adding the $\mathcal{O}_{\chi}$ deformation to the action,
the matrix elements we discussed above become
\begin{align}
  \langle n \pm 1 | e^{-\beta (H + \Delta H_{\chi})} | n \rangle 
  &=  \left(\beta L K_2 \Lambda^{2-\Delta_{\chi}} \mu^{\Delta_{\chi}}\, e^{-S_{2} \mp  2 i \frac{\theta}{N}}\right)^{(N-1)/2}\,,
\label{eq:NN_chi_odd}\\
\langle n \pm 2  | e^{-\beta (H + \Delta H_{\chi}) } | n \rangle 
&= \beta L K_2 \Lambda^{2-\Delta_{\chi}} \mu^{\Delta_{\chi}}\, e^{-S_{2} \pm  2 i \frac{\theta}{N}}  \,.
\label{eq:NNN_chi_odd}
\end{align}
The reason the matrix element \eqref{eq:NN_chi_odd} is non-zero is that if one
starts in vacuum $n$, then tunneling to the  next-to-nearest-neighbor on the
`right' $(N-1)/2$ times puts one in the vacuum with label $n-1$, due to the
fact that the label $n$ has a periodicity of $N$.

We can summarize this by observing that for even $N$, there is no tunneling to
nearest neighbor vacua at all.  The holonomy effective potential vacua split
into two sets that never mix with each other under time evolution.  This is a
consequence of the mixed 't Hooft anomaly between the $\Z_2$ subgroup of the
$\Z_N$ $1$-form symmetry and the unbroken $\Z_2$ subgroup of the $0$-form chiral
symmetry.  However, for odd $N$, the amplitude to move to nearest neighbor
minima is non-zero:  it occurs at order $\frac{N-1}{2}$ in semiclassics, and is suppressed by
$e^{- \frac{N-1}{2} S_{2}}$. Of course, the disparity between even and odd $N$ disappears at large $N$.

Next we consider the behavior of the string tension.  In  the Born-Oppenheimer
approximation, the tunneling Hamiltonian can be written as
\begin{align} 
H_{\rm BO} = - \sum_{n=1}^{N} K_2 \Lambda^{2-\Delta_{\chi}} \mu^{\Delta_{\chi}} e^{-S_2 + 2 i \frac{\theta}{N} } | n+2\rangle \langle n| + {\rm h.c.}
\end{align}
For even $N$, this Hamiltonian decompose to two decoupled Hamiltonians, one with
a sum over $n \in {0,2, \ldots, N-2} $ and the other with a sum over $n \in
{1,3, \ldots, N-1} $. These sets of states remain unmixed  at any
non-perturbative order. For odd $N$, there is no such decomposition.  

\begin{figure}[t]
\vspace{-1.cm}
\begin{center}
\includegraphics[width = 0.7\textwidth]{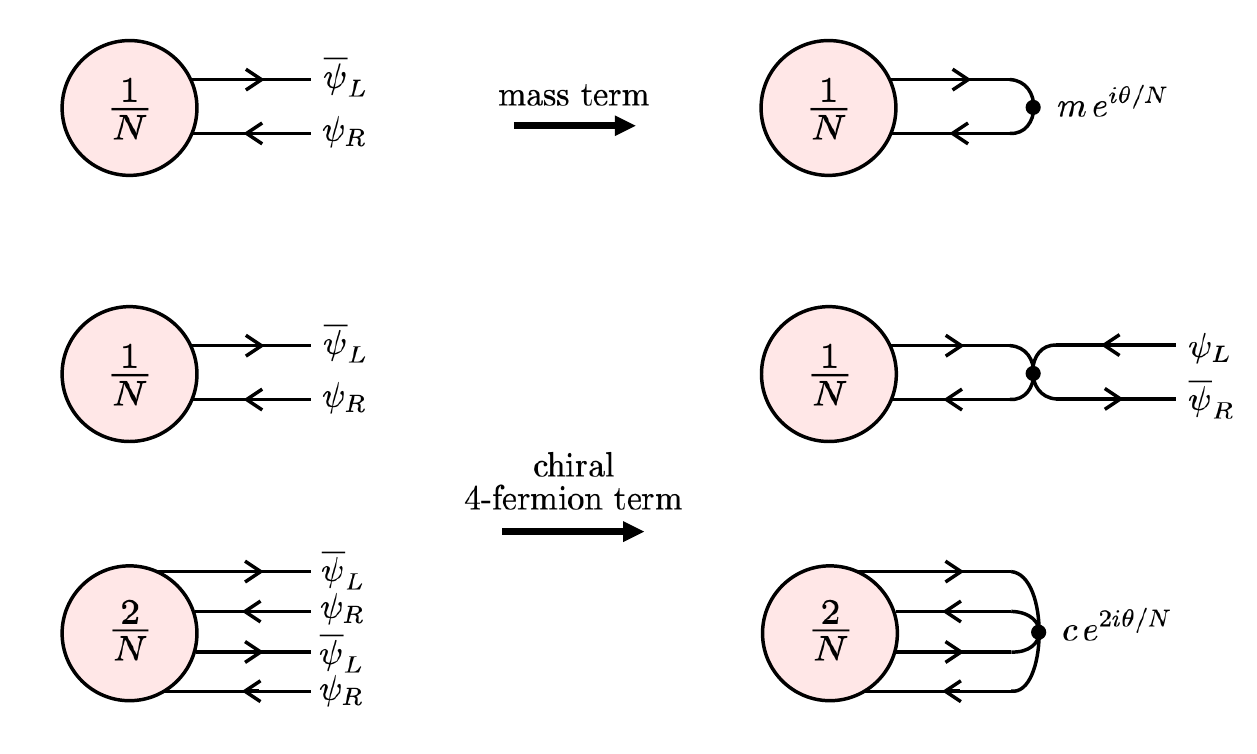}
\caption{(Top) In the semi-classical domain of massive Schwinger model,
confinement is generated by fractional instantons where zero modes are soaked up
by mass term. (Middle) In massless Schwinger model with four-fermion deformations, it
is not possible to lift the zero modes of a fractional instanton, but just flip
its chirality.  
(Bottom) But it is possible to lift up the zero modes of a fractional instanton
with topological charge $Q=2/N$. These configurations causes confinement for
external probe charges. }
\label{fig:topologicalconfig}
\end{center}
\end{figure}
 
The energy spectrum can be obtained by diagonalizing $H_{\rm BO}$ and is given by 
\begin{align}
   E_k(\theta) = -K_2 \Lambda^{2-\Delta_{\chi}}\mu^{\Delta_{\chi}}  \, e^{- S_2} \cos  \left(  \frac{ 2(\theta+ 2 \pi k)} {N}\right).
  \end{align} 
Therefore, using the correlator  \eqref{cor}, we can deduce that the string tensions are:
\begin{align}
T_k   =  -  K_2 \Lambda^{2-\Delta_{\chi}} \mu^{\Delta_{\chi}} \, e^{- S_2}    \left[   \cos    \left(  \frac{ 2 (\theta + 2\pi k)}{N}  \right) -  \cos  \left(  \frac{ 2 \theta }{N}  \right)  \right] + O(c^2) \,.
\end{align} 
This result also extrapolates to the result we obtained on $\R^2$ in the
decompactification limit.  As discussed in the previous section, for  $- \pi <
\theta <\pi$, it leads to confinement for all external charges for which  $k
\neq 0, N/2$ (mod $N$) and screening for $k =  0, N/2$ (mod $N$).

\section{Conclusions}
At first glance the charge-$N$ Schwinger model is an attractive solvable toy
model for questions about quark confinement and chiral symmetry breaking: it has
a $1$-form $\mathbb{Z}_N$ chiral symmetry as well as a (discrete) chiral
symmetry, just as 4d gauge theories with massless vector-like fermions.  But it
has long been known that the dynamics of this 2d model is radically different
from  4d gauge theories:  it does not confine fundamental-representation test
charges in the chiral (vanishing charged fermion mass) limit.  Indeed, by now it
is common wisdom that 2d gauge theories with massless fermions do not confine,
with both abelian examples like the charge-$N$ Schwinger model and non-abelian
examples such as 2d $SU(N)$ QCD with one Majorana fermion (see e.g.
\cite{Dalley:1992yy,Bhanot:1993xp,Demeterfi:1993rs,Lenz:1994du,Kutasov:1994xq,Gross:1995bp,
Gross:1997mx,Smilga:1998dh,Katz:2013qua,Komargodski:2020mxz,Smilga:2021zrw,Dempsey:2021xpf}), among
others~\cite{Delmastro:2021otj}.

In view of the very sharp difference of this behavior of 2d gauge theories with
massless fermions from naive expectations based on more familiar 4d examples, it
is important to understand what drives these differences in a precise way.  A
natural but naive guess based on the paragraph above is that this behavior is
driven by masslessness of the fermions, along with peculiarities of confinement
in 2d.  This is not quite correct.  Instead, the crucial issue\footnote{At
least in the absence of deformations of the action by local topological
operators --- such deformations make the story more
complicated, see Ref.~\cite{Cherman:2021nox} for an extensive discussion.} is
the presence (or absence) of appropriate mixed 't Hooft anomalies involving the
$1$-form symmetry. When appropriate 't Hooft anomalies are present the 2d gauge
theories lie in deconfined phases on $\mathbb{R}^2$, while when appropriate 't
Hooft anomalies are absent, they confine.

Our discussion in this paper gives a sharp illustration of this point 
in the context of the charge-$N$ Schwinger model.  We have analyzed a modified
version of the Schwinger model with dynamics that are closer to those of 4d
gauge theories. The standard massless Schwinger model has a mixed 't Hooft
anomaly between the $\Z_N$ $1$-form symmetry and the $\Z_N$ chiral symmetry,
which leads to deconfinement. Our modification involves turning on deformations of the action by two
four-fermion operators $\mathcal{O}_{\rm jj}$ and $\mathcal{O}_{\chi}$.  The $\mathcal{O}_{\rm jj}$ operator
is neutral under all of the symmetries of the model, while $\mathcal{O}_{\chi}$ has charge
$2$ under chiral symmetry.  The basic idea is that when it is relevant, the
$\mathcal{O}_{\chi}$ deformation reduces the chiral symmetry. However, since $\mathcal{O}_{\chi}$
has chiral charge $2$, it can preserve a $\mathbb{Z}_2$ chiral symmetry, and the
charged fermions do not get a mass term.\footnote{This assumes that we use a regulator that preserves $\Z_2$ chiral symmetry.} At the same time, the 't Hooft anomaly
structure is altered, and so the theory should confine.

Let us review this in a little bit more detail.   The mass operator has chiral
charge $1$. When $N$ is even, and the story is simplest,  the effect of these
deformations is to break the chiral symmetry from $\mathbb{Z}_N$ to
$\mathbb{Z}_2$ when $N$ is even. Provided the coefficient of $\mathcal{O}_{\rm jj}$ is
positive and large enough, the $\mathcal{O}_{\chi}$ operator is relevant, and so the
chiral symmetry remains $\mathbb{Z}_2$ deep in the infared.  This symmetry is
enough to forbid the fermion mass term, so the deformed theory must be viewed as
a variant of the massless Schwinger model.  We analyzed the behavior of this
theory on $\mathbb{R}^2$ and $\mathbb{R}\times S^1$, and showed that it confines
fundamental test charges when $N$ is even and is larger than $2$.  Test charges
with charge $q = N/2$ remain deconfined to a residual mixed 't Hooft anomaly
between the $\Z_2$ subgroup of the $1$-form symmetry and the unbroken chiral
symmetry.  
When $N$ is odd, the $\mathcal{O}_{\chi}$ perturbation breaks chiral symmetry completely.  Depending on whether it is relevant or irrelevant, it may or may not be possible to get an emergent chiral symmetry in the infrared, see Footnote~\ref{footnote:Wilson}.  When there is an emergent chiral symmetry in the infrared, the odd $N$ model remains deconfined even with $\mathcal{O}_{\chi}$.   Of course, when the $\mathcal{O}_{\chi}$ operator is relevant and $N$ is odd,  test charges of all
representations with non-zero $N$-ality are confined. 

Our results on relevance versus irrelevance of the operator $\mathcal{O}_{\chi}$ follow
from standard properties of the abelian bosonization dictionary.  We reached our
results on confinement in three basic ways. First, they essentially follow from
the totalitarian principle of QFT combined with symmetry arguments:  in the
absence of an 't Hooft anomaly forcing deconfinement, and without fine-tuning of
relative vaccuum energies of universes in this model (which could be done using
deformations by local topological operators without breaking any additional
symmetries), nothing forbids an area law for large Wilson loops, so one-form
symmetries should not be spontaneously broken in $1+1$d QFTs. Second, we took
advantage of bosonization to explicitly analyze the behavior of the theory on
$\mathbb{R}^2$. For suitable values of deformation parameters bosonization
allows one to systematically calculate observables like string tensions in
abelian 2d gauge theories, which would normally require an intractable
strong-coupling analysis in more complicated theories. We then examined the
behavior of theory on $\R \times S^1$ for small $S^1$ (compared to e.g.
$1/(eN)$).  This has the advantage that one can use either the bosonized
or fermionic duality frames, and allowed us to explore the confinement mechanism
in some detail. Confinement is driven by instantons with fractional topological
charge, with a number of interesting parallels to confinement in 4d gauge
theories on $\R^3 \times S^1$.  The instanton physics in this model also has
some unusual features: for example, there are finite-action configurations with zero
topological charge with robust fermion zero modes.

Much of this has interesting parallels to $SU(N)$  QCD with one adjoint Majorana
fermion in 2d, see \cite{Gross:1995bp,Cherman:2019hbq}. It was pointed out in
Ref.~\cite{Cherman:2019hbq} that 2d adjoint QCD also has two four-fermion
deformations which preserve all of the standard symmetries of the model.  In
contrast to the 2d Schwinger model, the four-fermion deformations of 2d adjoint
QCD are both classically marginal. Adjoint QCD in 2d can also be deformed by adding
local topological operators to the action without breaking any standard
symmetries~\cite{Cherman:2021nox}. At generic points in the parameter space of
the model, there's no way to forbid an area law for large Wilson loops (the 't
Hooft anomalies are not rich enough to do this in general), and so the model
should confine at generic points in this parameter space.  Note that this is a
statement about the theory with the fermion mass term set to zero, where it is
protected by symmetries from radiative corrections.  
Nevertheless, there is a corner in this parameter space where the
model \emph{does} deconfine, as explained in Ref.~\cite{Komargodski:2020mxz}.
In this corner of the parameter space, there is an extra unconventional
symmetry which is generated by non-invertible topological line
operators.  These non-invertible lines are charged under the $1$-form
symmetry, which means that they participate in a mixed 't Hooft anomaly.
This can be used to show that the area law term cannot arise in
expectation values of large 't Hooft loops.  However, just as in the
charge-$N$ Schwinger model, the existence of chiral-symmetry-preserving deformations that drive the
theory to confine means that keeping the charged fermion mass term set
to zero is not \emph{by itself} enough to drive deconfinement in 2d gauge theories.

\acknowledgments
We are very grateful to Zohar Komargodski and Yuya Tanizaki  for early discussions on the deformations of the Schwinger model and to Alexei Tsvelik for explanations of some issues in bozonization and scaling.
M.S. is supposed in part by DOE Grant No. DE-SC0011842. The work of 
M.U. is supported by U.S. Department of Energy,
Office of Science, Office of Nuclear Physics under Award Number DE-FG02-03ER41260. M.S. and A.V. are grateful to Kavli Institute for Theoretical Physics where their work was supported 
by the National Science Foundation under Grant No. NSF PHY-1748958.
A.C. and T.J. thank the Simons Center for Geometry and Physics for hospitality during the completion of this work.

\appendix

\section{Bosonization and holonomy effective potential}\label{sec:appendix}

In this appendix we use bosonization to compute the exact effective
potential for the gauge field holonomy in the massless Schwinger model.
This amounts to solving the theory on a torus. We emphasize certain
aspects of the global structure of the bosonized theory that are
particularly important in obtaining the correct result for the holonomy
effective potential.  A closely related discussion of axion-like
couplings for abelian gauge fields appears in Ref.~\cite{Cordova:2019jnf}.

We work on $T^2 = S^1_L\times S^1_\beta$, with coordinates $(\tau,x)\sim (\tau+\b,x+L)$. The gauge field satisfies the periodicity conditions
\begin{align}
a(\t+ \b, x) = a(\t,x) + dh_\t(\t,x)\,, \quad a(\t,x+L) = a(\t,x) + dh_x(\t,x), \label{eq:bcs}
\end{align}
where $h_\t,h_x$ are transition functions subject to the consistency condition
\begin{equation} 
dh_\t(\t,x+L)+dh_x(\t,x)  = dh_\t(\t,x) + dh_x(\t+\b,x)\,. \label{eq:cocycle}
\end{equation}
In a $U(1)$ gauge theory, we demand that the transition functions satisfy the cocycle condition
\begin{align}
h_\t(\t,x+L) - h_\t(\t,x) - h_x(\t+\b,x) + h_x(\t,x) = 2\pi Q,
\end{align}
where $Q \in\mathbb{Z}$. The integer $Q$ is in fact the topological charge on the torus,
\begin{align}
\frac{1}{2\pi}\int_{T^2} da &= 2\pi Q\,.
\end{align}
In the path integral we sum over transition functions satisfying the cocycle condition for fixed $Q$, and sum over $Q$. Gauge transformations act as $a \to a + d\lambda$ where $\lambda(\t,x)$ is an arbitrary real-valued function.\footnote{We are used to requiring that gauge transformations are single-valued on the torus mod $2\pi$. In the current presentation, this is an unnecessary assumption. If we add charge-$1$ matter, for instance, it obeys $\phi(\t+\b,x) = e^{ih_\t(\t,x)}\phi(\t,x), \, \phi(\t,x+L) = e^{ih_x(\t,x)}\phi(\t,x)$. These boundary conditions are gauge-invariant with respect to \eqref{eq:gaugetransformations}, even for gauge functions which are not single-valued on the torus. } The transition functions themselves transform under gauge transformations (redundancies) as
\begin{align} \label{eq:gaugetransformations} 
h_{\t}(\t,x) \to h_{\t}(\t,x) + \lambda(\t+\b,x)-\lambda(\t,x) + 2\pi m_\t\,, \\[1mm]
 h_{x}(\t,x) \to h_{x}(\t,x) + \lambda(\t,x+L)-\lambda(\t,x) + 2\pi m_x\,,
\end{align}
which leaves the cocycle condition and the integer $Q$ invariant. 

The compact scalar $\varphi$ can wind around the two cycles of the torus
\begin{align}
\varphi(\t+\b,x) = \varphi(\t,x) + 2\pi n_\t\,, \quad \varphi(\t,x+L) = \varphi(\t,x) + 2\pi n_x\,. \label{eq:winding} 
\end{align}
Naively, the bosonization rules suggest that the coupling between $\varphi$ and $a$ to be 
\begin{align}
\frac{iN}{2\pi} \int_{\t_\star}^{\t_\star +\beta}d\t\int_{x_\star}^{x_\star+L}dx \, a_\mu(\t,x)\epsilon^{\mu\nu}\partial_\nu\varphi(\t,x)\,.
\end{align} 
Here we have chosen an arbitrary basepoint $(\t_\star,x_\star)$ of the torus. The above integral suffers from two problems: it is not gauge invariant mod $2\pi i$, and it depends explicitly on the choice of reference point.\footnote{The fact that directly using the bosonization rules gives rise to gauge non-invariant terms in the Schwinger model on compact spacetimes was noticed in Ref.~\cite{PhysRevD.54.7757}. } Following Refs.~\cite{Cordova:2019jnf,PhysRevB.103.195113}, we remedy these issues by including correction terms involving the transition functions and topological charge:
\begin{align}\label{eq:extra}
``\frac{i}{2\pi}\int a\wedge d\varphi " &=\frac{i}{2\pi}\int_{\t_\star}^{\t_\star+\beta}d\t\int_{x_\star}^{x_\star+L}dx\, a_\mu\epsilon^{\mu\nu}\partial_\nu\varphi \nonumber\\
&-\frac{i}{2\pi}\int_{x_\star}^{x_\star+L}dx\, h_\t(\t_\star,x)\partial_x\varphi(\t_\star,x) + \frac{i}{2\pi}\int_{\t_\star}^{\t_\star+\beta}d\t\, h_x(\t,x_\star)\partial_\t\varphi(\t,x_\star) \nonumber\\
&+i Q\varphi(\t_\star,x_\star)\,.
\end{align}
One can readily verify that the above result is manifestly gauge invariant, and (by repeated use of the cocycle condition) that it is independent of the choice of $\t_\star,x_\star$. We may also integrate by parts to find a consistent expression for the typical `BF' form of the topological coupling, 
\begin{align}
``\frac{i}{2\pi}\int \varphi\wedge da " &= \frac{i}{2\pi}\int_{\t_\star}^{\t_\star+\beta}d\t\int_{x_\star}^{x_\star+L}dx\, \varphi(\partial_\t a_x-\partial_x a_\t)\nonumber \\
&+ in_\t \left[h_x(\t_\star,x_\star)-\int_{x_\star}^{x_\star+\beta}dx\,a_x(\t_\star,x)\right]\nonumber\\
&-i n_x\left[h_\t(\t_\star,x_\star) - \int_{\t_\star}^{\t_\star+\beta}d\t\, a_\t(\t,x_\star)\right] \,.
\end{align}
The quantities in brackets in the above correction terms are the (gauge-invariant) holonomies of the gauge field around the two non-contractible cycles. 

\subsection{Dimensional reduction}

Taking inspiration from Ref.~\cite{Misumi:2019dwq} we now consider the dimensional reduction
of the theory on a circle, taking $eL \ll 1$. We start with the usual
expression for the bosonized action, 
\begin{equation}
\mathcal L \overset{!}{=} \frac{1}{2e^2}|da|^2 + \frac{R^2}{2}|d\varphi|^2 + \frac{i N}{2\pi}a\wedge d\varphi\,,
\end{equation}
and consider the correction terms at the end. We take Coulomb gauge $\partial_xa_x = 0$. This implies that the transition functions satisfy $\partial_x^2h_\t = 0$ and $\partial_x h_x = 0$. We have further gauge freedom with gauge functions satisfying $\partial_x^2\lambda =0$, which we can use to set $h_x = 0$, while the cocycle condition fixes
\begin{equation}
h_\t(\t,x) = \frac{2\pi Q\, x}{L} + g(\t)
\end{equation}
for some arbitrary real-valued function $g(\t)$. Hence, the boundary conditions on the gauge field are
\begin{align}
a_\t(\t+\b,x) = a_\t(\t,x) + \partial_\t g(\t),\quad a_x(\t+\b) = a_x(\t) + \frac{2\pi Q}{L}\,, 
\end{align}
and $a_\t, a_x$ are periodic in the $x$ direction. Varying $a_\t,$ we find the Gauss law
\begin{equation}
\partial_x^2 a_\tau = \frac{i Ne^2}{2\pi}\partial_x\varphi\,.
\end{equation}
If we integrate both sides over $x$, we find that $\varphi$ cannot wind around $S^1_L$. This sets $n_x =0$ in Eq.~\eqref{eq:winding}, so that without loss of generality we can make the decomposition
\begin{align}
\varphi(\t,x) &= \sum_{\ell\in\mathbb{Z}} \varphi_\ell(\tau)e^{2\pi i \ell (x-x_\star)/L}, \nonumber \\
a_\t(\t,x) &= \sum_{\ell\in\mathbb{Z}}a_{\t,\ell}(\t)e^{2\pi i \ell (x-x_\star)/L},
\end{align} 
and the Gauss law gives the relation $a_{\t,\ell}(\t) = \frac{Ne^2L}{4\pi^2\ell}\,\varphi_\ell(\t)$ (for the nonzero modes). Plugging in the Kaluza-Klein decomposition and integrating over $x$ gives the naive form of the effective action on $S^1_\beta$, 
\begin{align}
S \overset{!}{=} L \int_{\t_\star}^{\t_\star+\beta}d\t\,\Bigg\{ &\frac{1}{2e^2}(\partial_\t a_x)^2 + \frac{R^2}{2}(\partial_\t\varphi_0)^2-\frac{iN}{2\pi}\,a_x\partial_\t\varphi_0 \nonumber \\
&+R^2\sum_{\ell>0}\left[\partial_\t\varphi_\ell\partial_\t\varphi_{-\ell} + \left(\left(\frac{2\pi \ell}{L}\right)^2 +\frac{N^2e^2}{4\pi^2R^2}\right)\varphi_\ell \varphi_{-\ell}  \right] \Bigg\}
\end{align}
The correction terms from Eq.~\eqref{eq:extra} give
\begin{align}
i NQ\varphi(\t_\star,x_\star)&-\frac{iN}{2\pi}\int_{x_\star}^{x_\star+L}dx\, \left[\frac{2\pi Q\, x}{L}+g(\t_\star)\right]\partial_x\varphi(\t_\star,x) \nonumber\\
&=i NQ\varphi(\t_\star,x_\star)-\frac{iN}{2\pi}2\pi Q\left[\varphi(\t_\star,x_\star)-\varphi_0(\t_\star)\right] = i N Q\, \varphi_0(\t_\star). 
\end{align}
Therefore, ignoring the non-zero modes of $\varphi$, which decouple exactly, we are left with 
\begin{align}
S = i N Q\, \varphi_0(\t_\star) + L \int_{\t_\star}^{\t_\star+\beta}d\t\,\left[ \frac{1}{2e^2}(\partial_\t a_x)^2 + \frac{R^2}{2}(\partial_\t\varphi_0)^2-\frac{iN}{2\pi}\,a_x\partial_\t\varphi_0\right]\,. \label{eq:circle_action}
\end{align}
The additional term involving the topological charge, which descended from the construction on $T^2$, matches the same term used in Ref.~\cite{doi:10.1063/5.0060808} to properly define the Lagrangian describing the above quantum mechanical theory of $N$ degenerate states. 

\subsection{Holonomy potential}
\label{sec:GPY_derivation}
We are now in a position to derive the holonomy effective potential. Recall that in our chosen gauge, the holonomy around $S^1_L$ is
\begin{align}
e^{-ih_x(\t,x_\star)}e^{i \int_{x_\star}^{x_\star+L}dx\, a_x(\t,x)} = e^{i L a_x(\t)} \equiv e^{i a(\t)}. 
\end{align}
For convenience, we drop the subscript on $\varphi_0$, choose $\t_\star = 0$, and replace $n_\t \to P$. The boundary conditions obeyed by the fields are
\begin{equation}
a(\t+\beta) = a(\t) + 2\pi Q, \quad \varphi(\t+\beta) = \varphi(\t) + 2\pi P,
\end{equation}
with the values of $Q,P$ summed over in the path integral. For completeness, we derive the holonomy effective potential in the presence of an insertion $e^{i \ell \varphi(\t_0)}$. In order to get a non-vanishing result we have to modify the boundary conditions to 
\begin{equation}
a(\t+\beta) = a(\t) + 2\pi \left(Q+\frac{\ell}{N}\right). 
\end{equation}
The action in that case is
\begin{equation} \label{action}
S =\int_0^\beta d\t\,\left[ \frac{1}{2e^2L}\dot a^2 + \frac{LR^2}{2}\dot \varphi^2 - \frac{iN}{2\pi}a\,\dot\varphi \right] +i N (Q+\tfrac{\ell}{N}) \, \varphi(0) -i\ell\,\varphi(\t_0). 
\end{equation}
Note that we had to modify the correction term to account for the new boundary conditions. We perform a mode decomposition consistent with the boundary conditions, 
\begin{equation}
\varphi(\t) =\frac{2\pi P}{\beta}\t+  \sum_{k \in \mathbb Z} \varphi_k \, e^{i\frac{2\pi k}{\beta} \t}\, \quad a(\t) = \frac{2\pi \left(Q+\frac{\ell}{N}\right)}{\beta}\t + \sum_{k \in \mathbb Z} a_k\, e^{i\frac{2\pi k}{\beta} \t}.
\end{equation}
The various terms in the action become, after completing the square, 
\begin{align}
S &= \frac{2\pi^2}{e^2L\beta}(Q+\tfrac{\ell}{N})^2 +\sum_{k\not=0} \frac{2\pi^2k^2}{e^2L\beta}|a_k|^2 + \frac{2\pi^2LR^2}{\beta}P^2\nonumber \\
&-iN\pi (Q+\tfrac{\ell}{N})P-iNP(a_0+\tfrac{2\pi\ell}{N}\tfrac{\t_0}{\beta}) +iNQ \, \varphi_0 \nonumber \\
&+ \sum_{k\not=0} \frac{\beta N^2}{8\pi^2LR^2} \left|a_k+\tfrac{i\ell\, \omega^{-k}}{kN}\right|^2  - \frac{2\pi^2LR^2}{\beta} \sum_{k\not=0}k^2\left|\varphi_k - \frac{\beta N}{4\pi^2LR^2}\frac{(a_k+\tfrac{i\ell\,\omega^{-k}}{kN})}{k}\right|^2\,.
\end{align}
where we defined $\omega = e^{2\pi i \t_0/\beta}$ and we have used the fact that $a(t)$ is real so $a_{-k} = a_k^*$. Integrating over $\varphi_0$ sets $Q=0$, and integrating over $\varphi_k$ for $k\not=0$ gives an overall holonomy-independent constant. Poisson resummation on $P$ gives
\begin{align}
\sum_{P\in\Z}e^{-\frac{2\pi^2LR^2}{\beta}P^2-iNP(a_0+\tfrac{\pi\ell}{N}+\tfrac{2\pi\ell}{N}\tfrac{\t_0}{\beta})} = \sqrt{\frac{\beta}{2\pi LR^2}}\,\sum_{n\in\Z}\, e^{-\frac{\beta  N^2}{8\pi^2LR^2}\left(a_0-\tfrac{2\pi\ell}{N}(\tfrac{1}{2}-\tfrac{\t_0}{\beta})-\frac{2\pi n}{N}\right)^2}\,.
\end{align}
Dropping the overall multiplicative factor, we have
\begin{align}
S_{\text{eff},n} &= \beta \frac{2\pi^2}{e^2L\beta^2}\left(\frac{\ell}{N}\right)^2 + \beta\sum_{k\not=0}\frac{1}{2e^2L}\left(\frac{2\pi k}{\beta}\right)^2|a_k|^2 \nonumber\\
&+ \beta \frac{N^2}{8\pi^2LR^2}\left(a_0-\frac{2\pi\ell}{N}(\tfrac{1}{2}-\tfrac{\t_0}{\beta})-\frac{2\pi n}{N}\right)^2 + \beta\sum_{k\not=0} \frac{ N^2}{8\pi^2LR^2} \left|a_k+\tfrac{i\ell\, \omega^{-k}}{kN}\right|^2\,. 
\end{align}
Noting that the Heaviside step function $\Theta$ can be written as
\begin{equation}
\Theta(\t-\t_0) = \frac{1}{2}+\frac{\t-\t_0}{\beta}-\frac{i}{2\pi}\sum_{k\not=0}\frac{1}{k}\, e^{2\pi i k(\t-\t_0)/\beta}\, , 
\end{equation}
the above expression is equivalent to (using $m_\gamma = \frac{eN}{2\pi R}$)
\begin{align}\label{eq:appendix_GPY}
S_{\text{eff},n} = \int_0^\beta d\t\, \left[ \frac{1}{2e^2L}\dot a^2 + \frac{m_\gamma^2}{2e^2L}\left(a-\frac{2\pi n}{N}-\frac{2\pi\ell}{N}\Theta(\t-\t_0)\right)^2 \right]\,,
\end{align}
where the function $a$ satisfies $a(\beta) = a(0)+\tfrac{2\pi\ell}{N}$. This gives Eq.~\eqref{eq:GPY_insertion} in the main text. 

\section{Scaling dimensions}
\label{sec:scaling_dim}
In this appendix we calculate the scaling dimensions of operators as functions of the marginal parameter $g$, which is the coefficient of $\mathcal{O}_{jj}$ operator, by summing some simple classes of diagrams in the fermionic presentation of our model.  This reproduces the results from our analysis using bosonization in the main text.  The fact that just summing these simple classes of diagrams already reproduces the exact results from bosonization means that all of the  Feynman diagrams that we did not consider conspire to cancel, but showing this explicitly is a non-trivial project which did not manage to finish. Given the well-established nature of the bosonization approach, we do not pursue this more ambitious project here.  Before coming to the calculations, we note that we will use Minkowski notation in this Appendix, apart from using Euclidean rotations to evaluate the final momentum integrals.

Let us first consider the normalization in the relation \eqref{eq:current_map} for bozonization of the current $j_\mu=\bar \psi\gamma_\mu \psi$. So called Bjorken limit is very instrumental in this respect. For two operators, $A$ and $B$, the limit $q_0\to \infty$ in $\int d^2 x\, e^{iqx} \mathrm{T}\{A(x) B(0)\}$ gives the equal-time commutator of $A$ and $B$,
\begin{equation}
  \lim_{q_0\to \infty} q_0 \int d^2 x\, e^{iqx} \mathrm{T}\{A(x) B(0)\}=i\int d^2x\, e^{iqx}\,[A(x),B(0)]\,\delta(x_0)\,.
\end{equation}
\begin{figure}[h]
\begin{center}
\includegraphics[width = 0.7\textwidth]{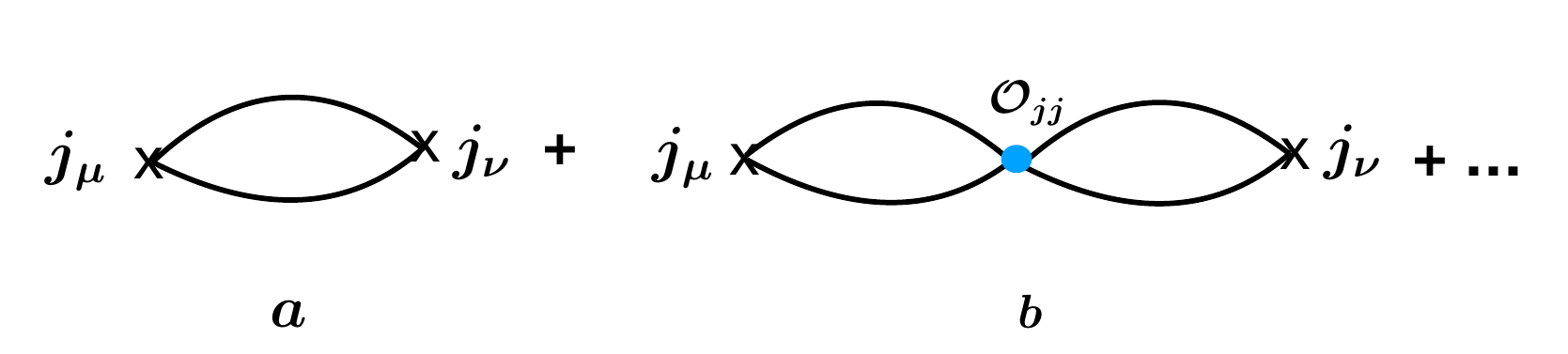}
\vspace{-.5cm}
\caption{Corrections to polarization due to $\mathcal{O}_{jj}$. }
\label{fig:g_correction}
\end{center}
\end{figure}
When operators $A$ and $B$ are components of the vector current $j^\mu=\bar \psi \gamma^\mu \psi$, the leading contribution to this correlation function ( a polarization operator) is given by the diagram $a$ in Fig.\,\ref{fig:g_correction}:
\begin{equation}
-i\Pi_{\mu\nu}(q)=\int d^2 x\, e^{iqx} \langle\,\mathrm{T}\{j_\mu(x) j_\nu(0)\}\rangle=\frac{i}{\pi q^2}\,(q_\mu q_\nu -g_{\mu\nu}q^2)\,.
\label{eq:scaling_dim_2}
\end{equation}
In $q_0\to \infty$ limit, this defines the $c$-number part of $[j_0(x),j_1(0)]\delta(x_0)$ commutator,
\begin{equation}
  \int d^2x e^{iqx} [j_0(x),j_1(0)]\delta(x_0) =\frac{q_1}{\pi}\,,
\end{equation}
and fixes the coefficient in bozonization of the current $j_\mu$,
\begin{equation}
  j_\mu \Longrightarrow \,-\frac{1}{2\pi}\,\epsilon_{\mu\nu}\partial^\nu \phi\,.
\end{equation}
Of course, the same result \eqref{eq:scaling_dim_2} for the polarization operator arises from the bosonic form of the current when the kinetic term of the $\phi$
 field is $(1/8\pi)\partial_\mu \phi \partial^\mu \phi$.
 
Switching on the $\mathcal{O}_{jj}$ operator leads to appearance of the second and higher loops for the correlator \eqref{eq:scaling_dim_2}.
In the first order in $g$ it is given by the two-loop diagram $b$ in Fig.\,\ref{fig:g_correction}. This is  a product, 
$(-2g)\Pi_{\mu\gamma}\Pi^\gamma_\nu=(-2g/\pi)\Pi_{\mu\nu}$,  of two one-loop polarization operators resulting in the factor $(-2g/\pi)$. Accounting for higher loops gives a geometrical progression so we come to overall factor $1/(1+2g/\pi)$,  consistent with bosonic considerations where the kinetic term becomes $(R^2/2)\,
\partial_\mu \phi \partial^\mu \phi$ with $R^2=(1+2g/\pi)/4\pi$\,.

\begin{figure}[h]
\begin{center}
\includegraphics[width = 0.7\textwidth]{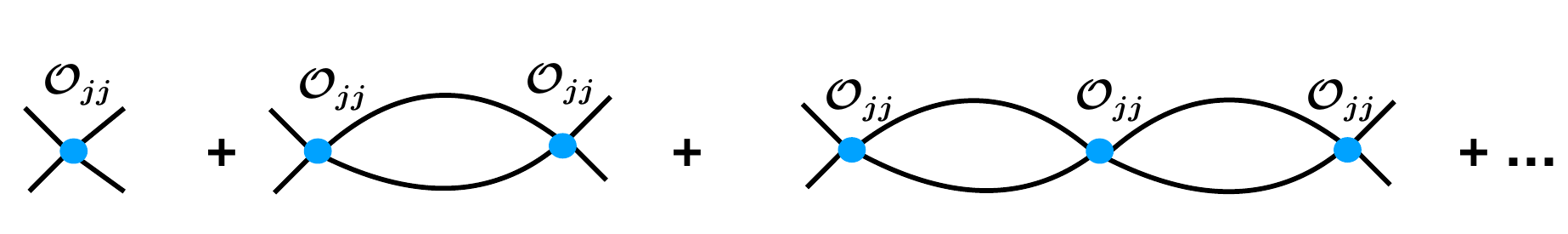}
\vspace{-.3cm}
\caption{Iterations of the operator $\mathcal{O}_{jj}$. }
\label{fig:Og_corrections}
\end{center}
\end{figure}

This phenomenon of finite renormalization of the current also shows up for the operator $\mathcal{O}_{jj}$. Iterations of this operator are illustrated by diagrams in Fig.\,\ref{fig:Og_corrections}\,.
 The same geometrical progression appears what could be interpreted as the effective substitution for the coupling $g$,
 \begin{equation}
  g \Longrightarrow \,\frac{g}{1+2g/\pi}\,.
\label{eq:g_substitution}  
\end{equation}

A word of caution should be added here. The above consideration of loop corrections for the operator $\mathcal{O}_{jj}$ refers to the pairing of bilinear operators $j_\mu=\bar \psi \gamma_\mu \psi$
in the loop. However, the loop corrections for  $\mathcal{O}_{jj}$ also include other bilinears,
namely, $\bar \psi_L \psi_R$, $\psi_L \psi_L$, $\psi_R \psi_R$ paired with corresponding Hermitian conjugated ones. The individual loops are logarithmicaly divergent, so it could lead to breaking of the marginal nature of  the operator $\mathcal{O}_{jj}$. Interestingly enough there is a cancellation between channels with fermion charge 0, like $\bar \psi_L \psi_R$, and double fermion charge, like 
$\psi_L \psi_L$. Altogether, the marginality of $\mathcal{O}_{jj}$ is preserved.\footnote{\,We are thankful to A.\,Tsvelik for for explaining this phenomenon to us.}

Let us consider now  bosonization of
 the operator $\bar\psi_L(x) \psi_R(x)$\,. Its scaling dimension is clearly equal to  1 for free fermions. Bosonization relates this operator to $e^{i\phi(x)}$\,.
To see that this bosonic operator has the same scaling dimension one can calculate the tadpole graphs. To this end let us start with one tadpole loop. 
\begin{align}\label{eq:scaling_dim1}
&e^{i\phi}=\sum_{n=0}^\infty \frac{1}{n!}\,(i\phi)^n\,;\quad 
(i\phi)^n \Longrightarrow \frac{n!}{2(n-2)!}(i\phi)^{n-2}\langle \,(i\phi)(i\phi)\,\rangle\,; \nonumber\\
&e^{i\phi}\Longrightarrow e^{i\phi}\, \frac{1}{2}\langle \,(i\phi)(i\phi)\,\rangle\,.
\end{align}
Here $\langle \,(i\phi)(x)(i\phi)(x)\,\rangle$ is the propagator of the field $i\phi$ from the point $x$ to the same point. In Euclidean momentum space
\begin{equation}
  \langle \,(i\phi)(x)(i\phi)(x)\,\rangle =-\int \frac{d^2k}{(2\pi)^2}\,\frac{4\pi}{k^2}=-2\int_{\mu_1}^{\mu_2}\frac{dk}{k}=
  2\log\frac{\mu_1}{\mu_2}\,,
\end{equation}
where $\mu_1$ and $\mu_2$ are lower and upper cut-off in the momentum integration.
Proceeding in the same fashion with next tadpoles we get exponentiation of the log,
\begin{equation}
e^{i\phi}\Longrightarrow e^{i\phi}e^{\langle \,(i\phi)(i\phi)\,\rangle /2}=
\frac{\mu_1}{\mu_2}\,e^{i\phi}\,.
\end{equation}

Now let us switch on the operator $\mathcal{O}_{jj}$. On bosonic side it is just a change of 
the kinetic term coefficient and the corresponding change of  $\langle \,(i\phi)(x)(i\phi)(x)\,\rangle$ by the factor $1/(1+2g/\pi)$. This immediately shows that the scaling dimension instead of 1 becomes
\begin{equation}
    \Delta_{\,e^{i\phi}}=\frac{1}{1+{2g}/{\pi}}\,.
  \label{dim_phi}  
\end{equation}
On the fermion side one have to consider the loop diagram generated by the initial operator 
$\bar\psi_L(x) \psi_R(x)$ and the operator $\mathcal{O}_{jj}$. In the first order in $g$
\begin{align}
&\bar\psi_L(0) \psi_R(0)\,i\!\!\int d^2x\, (-g)\,\mathcal{O}_{jj}(x)=
4ig \,\bar\psi_L(0) \psi_R(0)\!\!\int d^2x\, \bar\psi_R(x) \psi_L(x)\bar\psi_L(x) \psi_R(x)\nonumber\\
&\Longrightarrow -4ig \int d^2x \,\langle \psi_L(x)\bar \psi_L(0)\rangle\,
\langle \psi_R(x)\bar \psi_R(0) \rangle \,\bar\psi_L(x) \psi_R(x)\,.
\end{align}
Here $\langle \psi_L(x)\bar \psi_L(0)\rangle$ and 
$\langle \psi_R(x)\bar \psi_R(0)\rangle$ denote fermionic propagators. In momentum space we come to 
\begin{equation}
 4ig \int \frac{d^2p}{(2\pi)^2}\,\frac{1}{p_R}\frac{1}{p_L}\,
   \bar\psi_L \psi_R=-\frac{2g}{\pi}\,\log\,\frac{\mu_1}{\mu_2}
   \;\bar\psi_L \psi_R
\end{equation}
Accounting for higher loops leads to two effects. First, it leads to the exponentiation of
the one loop result, and second, it leads to the substitution 
\eqref{eq:g_substitution} for the coupling $g$\,. Altogether, we get for the scaling dimension,
\begin{equation}
 \Delta_{\,\bar\psi_L\psi_R}=1-\frac{2g}{1+2g/\pi}=\frac{1}{1+2g/\pi}\,,
\end{equation}
where 1 comes from canonical dimension. This coincides with \eqref{dim_phi}.

Derivation of \eqref{eq:Ochi_dim} for the scaling dimension of the operator $\mathcal{O}_\chi$ on both bosonic and fermionic sides is similar the one given above for the $\bar\psi_L\psi_R$ operator.


\bibliographystyle{utphys}
\bibliography{small_circle} 

\providecommand{\href}[2]{#2}\begingroup\raggedright\begin{thebibliography}{10}

\bibitem{PhysRev.128.2425}
J.~Schwinger, ``Gauge invariance and mass. ii,''
  \href{http://dx.doi.org/10.1103/PhysRev.128.2425}{{\em Phys. Rev.} {\bfseries
  128} (Dec, 1962) 2425--2429}.
  \url{https://link.aps.org/doi/10.1103/PhysRev.128.2425}.

\bibitem{Coleman:1975pw}
S.~R. Coleman, R.~Jackiw, and L.~Susskind, ``{Charge Shielding and Quark
  Confinement in the Massive Schwinger Model},''
  \href{http://dx.doi.org/10.1016/0003-4916(75)90212-2}{{\em Annals Phys.}
  {\bfseries 93} (1975) 267}.

\bibitem{Coleman:1976uz}
S.~R. Coleman, ``{More About the Massive Schwinger Model},''
  \href{http://dx.doi.org/10.1016/0003-4916(76)90280-3}{{\em Annals Phys.}
  {\bfseries 101} (1976) 239}.

\bibitem{Anber:2018jdf}
M.~M. Anber and E.~Poppitz, ``{Anomaly matching, (axial) Schwinger models, and
  high-T super Yang-Mills domain walls},''
  \href{http://dx.doi.org/10.1007/JHEP09(2018)076}{{\em JHEP} {\bfseries 09}
  (2018) 076}, \href{http://arxiv.org/abs/1807.00093}{{\ttfamily
  arXiv:1807.00093 [hep-th]}}.

\bibitem{Anber:2018xek}
M.~M. Anber and E.~Poppitz, ``{Domain walls in high-T SU(N) super Yang-Mills
  theory and QCD(adj)},'' \href{http://dx.doi.org/10.1007/JHEP05(2019)151}{{\em
  JHEP} {\bfseries 05} (2019) 151},
  \href{http://arxiv.org/abs/1811.10642}{{\ttfamily arXiv:1811.10642
  [hep-th]}}.

\bibitem{Armoni:2018bga}
A.~Armoni and S.~Sugimoto, ``{Vacuum structure of charge k two-dimensional QED
  and dynamics of an anti D-string near an O1--plane},''
  \href{http://dx.doi.org/10.1007/JHEP03(2019)175}{{\em JHEP} {\bfseries 03}
  (2019) 175}, \href{http://arxiv.org/abs/1812.10064}{{\ttfamily
  arXiv:1812.10064 [hep-th]}}.

\bibitem{Misumi:2019dwq}
T.~Misumi, Y.~Tanizaki, and M.~{\"U}nsal, ``{Fractional $\theta$ angle, 't
  Hooft anomaly, and quantum instantons in charge-$q$ multi-flavor Schwinger
  model},'' \href{http://dx.doi.org/10.1007/JHEP07(2019)018}{{\em JHEP}
  {\bfseries 07} (2019) 018}, \href{http://arxiv.org/abs/1905.05781}{{\ttfamily
  arXiv:1905.05781 [hep-th]}}.

\bibitem{Komargodski:2020mxz}
Z.~Komargodski, K.~Ohmori, K.~Roumpedakis, and S.~Seifnashri, ``{Symmetries and
  strings of adjoint QCD$_{2}$},''
  \href{http://dx.doi.org/10.1007/JHEP03(2021)103}{{\em JHEP} {\bfseries 03}
  (2021) 103}, \href{http://arxiv.org/abs/2008.07567}{{\ttfamily
  arXiv:2008.07567 [hep-th]}}.

\bibitem{Cherman:2020cvw}
A.~Cherman and T.~Jacobson, ``{Lifetimes of near eternal false vacua},''
  \href{http://dx.doi.org/10.1103/PhysRevD.103.105012}{{\em Phys. Rev. D}
  {\bfseries 103} no.~10, (2021) 105012},
  \href{http://arxiv.org/abs/2012.10555}{{\ttfamily arXiv:2012.10555
  [hep-th]}}.

\bibitem{Cherman:2021nox}
A.~Cherman, T.~Jacobson, and M.~Neuzil, ``{Universal Deformations},''
  \href{http://arxiv.org/abs/2111.00078}{{\ttfamily arXiv:2111.00078
  [hep-th]}}.

\bibitem{Hansson:1994ep}
T.~H. Hansson, H.~B. Nielsen, and I.~Zahed, ``{QED with unequal charges: A
  Study of spontaneous Z$_n$ symmetry breaking},''
  \href{http://dx.doi.org/10.1016/0550-3213(95)00360-5}{{\em Nucl. Phys. B}
  {\bfseries 451} (1995) 162--176},
  \href{http://arxiv.org/abs/hep-ph/9405324}{{\ttfamily arXiv:hep-ph/9405324}}.
  [Erratum: Nucl.Phys.B 456, 757--757 (1995)].

\bibitem{Gaiotto:2014kfa}
D.~Gaiotto, A.~Kapustin, N.~Seiberg, and B.~Willett, ``{Generalized global
  symmetries},'' \href{http://dx.doi.org/10.1007/JHEP02(2015)172}{{\em JHEP}
  {\bfseries 02} (2015) 172},
\href{http://arxiv.org/abs/1412.5148}{{\ttfamily arXiv:1412.5148 [hep-th]}}.

\bibitem{Polyakov:1976fu}
A.~M. Polyakov, ``{Quark confinement and topology of gauge groups},''
\href{http://dx.doi.org/10.1016/0550-3213(77)90086-4}{{\em Nucl. Phys.}
  {\bfseries B120} (1977) 429--458}.

\bibitem{Gross:1980br}
D.~J. Gross, R.~D. Pisarski, and L.~G. Yaffe, ``{QCD and instantons at finite
  temperature},''
\href{http://dx.doi.org/10.1103/RevModPhys.53.43}{{\em Rev. Mod. Phys.}
  {\bfseries 53} (1981) 43}.

\bibitem{Thirring:1958in}
W.~E. Thirring, ``{A Soluble relativistic field theory?},''
  \href{http://dx.doi.org/10.1016/0003-4916(58)90015-0}{{\em Annals Phys.}
  {\bfseries 3} (1958) 91--112}.

\bibitem{Johnson:1961cs}
K.~Johnson, ``{Solution of the equations for the Green's functions of a
  two-dimensional relativistic field theory},''
  \href{http://dx.doi.org/10.1007/BF02731566}{{\em Nuovo Cim.} {\bfseries 20}
  (1961) 773--790}.

\bibitem{Mueller:1971mar}
A.~H. Mueller and T.~L. Trueman, ``{Anomalous short-distance behavior of
  quantum field theory - a massive Thirring model},''
  \href{http://dx.doi.org/10.1103/PhysRevD.4.1635}{{\em Phys. Rev. D}
  {\bfseries 4} (1971) 1635--1652}.

\bibitem{Gomes:1972yb}
M.~Gomes and J.~H. Lowenstein, ``{Asymptotic scale invariance in a massive
  Thirring model},'' \href{http://dx.doi.org/10.1016/0550-3213(72)90168-X}{{\em
  Nucl. Phys. B} {\bfseries 45} (1972) 252--266}.

\bibitem{Sulejmanpasic:2019ytl}
T.~Sulejmanpasic and C.~Gattringer, ``{Abelian gauge theories on the lattice:
  $\theta$-Terms and compact gauge theory with(out) monopoles},''
  \href{http://dx.doi.org/10.1016/j.nuclphysb.2019.114616}{{\em Nucl. Phys. B}
  {\bfseries 943} (2019) 114616},
  \href{http://arxiv.org/abs/1901.02637}{{\ttfamily arXiv:1901.02637
  [hep-lat]}}.

\bibitem{Gorantla:2021svj}
P.~Gorantla, H.~T. Lam, N.~Seiberg, and S.-H. Shao, ``{A modified Villain
  formulation of fractons and other exotic theories},''
  \href{http://dx.doi.org/10.1063/5.0060808}{{\em J. Math. Phys.} {\bfseries
  62} no.~10, (2021) 102301}, \href{http://arxiv.org/abs/2103.01257}{{\ttfamily
  arXiv:2103.01257 [cond-mat.str-el]}}.

\bibitem{Cherman:2019hbq}
A.~Cherman, T.~Jacobson, Y.~Tanizaki, and M.~\"Unsal, ``{Anomalies, a mod 2
  index, and dynamics of 2d adjoint QCD},''
  \href{http://dx.doi.org/10.21468/SciPostPhys.8.5.072}{{\em SciPost Phys.}
  {\bfseries 8} no.~5, (2020) 072},
  \href{http://arxiv.org/abs/1908.09858}{{\ttfamily arXiv:1908.09858
  [hep-th]}}.

\bibitem{Sachs:1991en}
I.~Sachs and A.~Wipf, ``{Finite temperature Schwinger model},'' {\em Helv.
  Phys. Acta} {\bfseries 65} (1992) 652--678,
  \href{http://arxiv.org/abs/1005.1822}{{\ttfamily arXiv:1005.1822 [hep-th]}}.

\bibitem{PhysRevD.11.2088}
S.~Coleman, ``Quantum sine-gordon equation as the massive thirring model,''
  \href{http://dx.doi.org/10.1103/PhysRevD.11.2088}{{\em Phys. Rev. D}
  {\bfseries 11} (Apr, 1975) 2088--2097}.
  \url{https://link.aps.org/doi/10.1103/PhysRevD.11.2088}.

\bibitem{Fradkin:2013sab}
E.~H. Fradkin, {\em {Field Theories of Condensed Matter Physics}}, vol.~82.
\newblock Cambridge Univ. Press, Cambridge, UK, 2, 2013.

\bibitem{Jayewardena:1988td}
C.~Jayewardena, ``{SCHWINGER MODEL ON S(2)},'' {\em Helv. Phys. Acta}
  {\bfseries 61} (1988) 636--711.

\bibitem{Smilga:1993sn}
A.~V. Smilga, ``{Instantons in Schwinger model},''
  \href{http://dx.doi.org/10.1103/PhysRevD.49.5480}{{\em Phys. Rev. D}
  {\bfseries 49} (1994) 5480--5490},
  \href{http://arxiv.org/abs/hep-th/9312110}{{\ttfamily arXiv:hep-th/9312110}}.

\bibitem{Shifman:1994ce}
M.~A. Shifman and A.~V. Smilga, ``{Fractons in twisted multiflavor Schwinger
  model},'' \href{http://dx.doi.org/10.1103/PhysRevD.50.7659}{{\em Phys. Rev.}
  {\bfseries D50} (1994) 7659--7672},
\href{http://arxiv.org/abs/hep-th/9407007}{{\ttfamily arXiv:hep-th/9407007
  [hep-th]}}.

\bibitem{Hetrick:1995wq}
J.~E. Hetrick, Y.~Hosotani, and S.~Iso, ``{The Massive multi - flavor Schwinger
  model},'' \href{http://dx.doi.org/10.1016/0370-2693(95)00310-H}{{\em Phys.
  Lett. B} {\bfseries 350} (1995) 92--102},
  \href{http://arxiv.org/abs/hep-th/9502113}{{\ttfamily arXiv:hep-th/9502113}}.

\bibitem{Smilga:1996dn}
A.~V. Smilga, ``{Two-dimensional instantons with bosonization and physics of
  adjoint QCD(2)},'' \href{http://dx.doi.org/10.1103/PhysRevD.54.7757}{{\em
  Phys. Rev. D} {\bfseries 54} (1996) 7757--7773},
  \href{http://arxiv.org/abs/hep-th/9607007}{{\ttfamily arXiv:hep-th/9607007}}.

\bibitem{Rodriguez:1996zj}
R.~Rodriguez and Y.~Hosotani, ``{Confinement and chiral condensates in 2-D QED
  with massive n flavor fermions},''
  \href{http://dx.doi.org/10.1016/0370-2693(96)00240-7}{{\em Phys. Lett. B}
  {\bfseries 375} (1996) 273--284},
  \href{http://arxiv.org/abs/hep-th/9602029}{{\ttfamily arXiv:hep-th/9602029}}.

\bibitem{Radozycki:2008zt}
T.~Radozycki, ``{Instantons and the infrared behavior of the fermion propagator
  in the Schwinger Model},''
  \href{http://dx.doi.org/10.1140/epjc/s10052-008-0622-6}{{\em Eur. Phys. J. C}
  {\bfseries 55} (2008) 509--516},
  \href{http://arxiv.org/abs/0801.4399}{{\ttfamily arXiv:0801.4399 [hep-th]}}.

\bibitem{Unsal:2020yeh}
M.~\"Unsal, ``{Strongly coupled QFT dynamics via TQFT coupling},''
  \href{http://arxiv.org/abs/2007.03880}{{\ttfamily arXiv:2007.03880
  [hep-th]}}.

\bibitem{Behtash:2018voa}
A.~Behtash, G.~V. Dunne, T.~Schaefer, T.~Sulejmanpasic, and M.~\"Unsal,
  ``{Critical Points at Infinity, Non-Gaussian Saddles, and Bions},''
  \href{http://dx.doi.org/10.1007/JHEP06(2018)068}{{\em JHEP} {\bfseries 06}
  (2018) 068}, \href{http://arxiv.org/abs/1803.11533}{{\ttfamily
  arXiv:1803.11533 [hep-th]}}.

\bibitem{Gaiotto:2017yup}
D.~Gaiotto, A.~Kapustin, Z.~Komargodski, and N.~Seiberg, ``{Theta, time
  reversal, and temperature},''
  \href{http://dx.doi.org/10.1007/JHEP05(2017)091}{{\em JHEP} {\bfseries 05}
  (2017) 091},
\href{http://arxiv.org/abs/1703.00501}{{\ttfamily arXiv:1703.00501 [hep-th]}}.

\bibitem{Dalley:1992yy}
S.~Dalley and I.~R. Klebanov, ``{String spectrum of (1+1)-dimensional large N
  QCD with adjoint matter},''
  \href{http://dx.doi.org/10.1103/PhysRevD.47.2517}{{\em Phys. Rev. D}
  {\bfseries 47} (1993) 2517--2527},
  \href{http://arxiv.org/abs/hep-th/9209049}{{\ttfamily arXiv:hep-th/9209049}}.

\bibitem{Bhanot:1993xp}
G.~Bhanot, K.~Demeterfi, and I.~R. Klebanov, ``{(1+1)-dimensional large N QCD
  coupled to adjoint fermions},''
  \href{http://dx.doi.org/10.1103/PhysRevD.48.4980}{{\em Phys. Rev. D}
  {\bfseries 48} (1993) 4980--4990},
  \href{http://arxiv.org/abs/hep-th/9307111}{{\ttfamily arXiv:hep-th/9307111}}.

\bibitem{Demeterfi:1993rs}
K.~Demeterfi, I.~R. Klebanov, and G.~Bhanot, ``{Glueball spectrum in a
  (1+1)-dimensional model for QCD},''
  \href{http://dx.doi.org/10.1016/0550-3213(94)90236-4}{{\em Nucl. Phys. B}
  {\bfseries 418} (1994) 15--29},
  \href{http://arxiv.org/abs/hep-th/9311015}{{\ttfamily arXiv:hep-th/9311015}}.

\bibitem{Lenz:1994du}
F.~Lenz, M.~A. Shifman, and M.~Thies, ``{Quantum mechanics of the vacuum state
  in two-dimensional QCD with adjoint fermions},''
  \href{http://dx.doi.org/10.1103/PhysRevD.51.7060}{{\em Phys. Rev. D}
  {\bfseries 51} (1995) 7060--7082},
  \href{http://arxiv.org/abs/hep-th/9412113}{{\ttfamily arXiv:hep-th/9412113}}.

\bibitem{Kutasov:1994xq}
D.~Kutasov and A.~Schwimmer, ``{Universality in two-dimensional gauge
  theory},'' \href{http://dx.doi.org/10.1016/0550-3213(95)00106-3}{{\em Nucl.
  Phys. B} {\bfseries 442} (1995) 447--460},
  \href{http://arxiv.org/abs/hep-th/9501024}{{\ttfamily arXiv:hep-th/9501024}}.

\bibitem{Gross:1995bp}
D.~J. Gross, I.~R. Klebanov, A.~V. Matytsin, and A.~V. Smilga, ``{Screening
  versus confinement in (1+1)-dimensions},''
  \href{http://dx.doi.org/10.1016/0550-3213(95)00655-9}{{\em Nucl. Phys. B}
  {\bfseries 461} (1996) 109--130},
  \href{http://arxiv.org/abs/hep-th/9511104}{{\ttfamily arXiv:hep-th/9511104}}.

\bibitem{Gross:1997mx}
D.~J. Gross, A.~Hashimoto, and I.~R. Klebanov, ``{The Spectrum of a large N
  gauge theory near transition from confinement to screening},''
  \href{http://dx.doi.org/10.1103/PhysRevD.57.6420}{{\em Phys. Rev. D}
  {\bfseries 57} (1998) 6420--6428},
  \href{http://arxiv.org/abs/hep-th/9710240}{{\ttfamily arXiv:hep-th/9710240}}.

\bibitem{Smilga:1998dh}
A.~V. Smilga, ``{QCD at $\theta \sim \pi$},''
  \href{http://dx.doi.org/10.1103/PhysRevD.59.114021}{{\em Phys. Rev. D}
  {\bfseries 59} (1999) 114021},
\href{http://arxiv.org/abs/hep-ph/9805214}{{\ttfamily arXiv:hep-ph/9805214
  [hep-ph]}}.

\bibitem{Katz:2013qua}
E.~Katz, G.~Marques~Tavares, and Y.~Xu, ``{Solving 2D QCD with an adjoint
  fermion analytically},''
  \href{http://dx.doi.org/10.1007/JHEP05(2014)143}{{\em JHEP} {\bfseries 05}
  (2014) 143}, \href{http://arxiv.org/abs/1308.4980}{{\ttfamily arXiv:1308.4980
  [hep-th]}}.

\bibitem{Smilga:2021zrw}
A.~V. Smilga, ``{A comment on instantons and their fermion zero modes in
  adjoint QCD\_2},''
  \href{http://dx.doi.org/10.21468/SciPostPhys.10.6.152}{{\em SciPost Phys.}
  {\bfseries 10} no.~6, (2021) 152},
  \href{http://arxiv.org/abs/2104.06266}{{\ttfamily arXiv:2104.06266
  [hep-th]}}.

\bibitem{Dempsey:2021xpf}
R.~Dempsey, I.~R. Klebanov, and S.~S. Pufu, ``{Exact symmetries and threshold
  states in two-dimensional models for QCD},''
  \href{http://dx.doi.org/10.1007/JHEP10(2021)096}{{\em JHEP} {\bfseries 10}
  (2021) 096}, \href{http://arxiv.org/abs/2101.05432}{{\ttfamily
  arXiv:2101.05432 [hep-th]}}.

\bibitem{Delmastro:2021otj}
D.~Delmastro, J.~Gomis, and M.~Yu, ``{Infrared phases of 2d QCD},''
  \href{http://arxiv.org/abs/2108.02202}{{\ttfamily arXiv:2108.02202
  [hep-th]}}.

\bibitem{Cordova:2019jnf}
C.~C\'ordova, D.~S. Freed, H.~T. Lam, and N.~Seiberg, ``{Anomalies in the Space
  of Coupling Constants and Their Dynamical Applications I},''
  \href{http://dx.doi.org/10.21468/SciPostPhys.8.1.001}{{\em SciPost Phys.}
  {\bfseries 8} no.~1, (2020) 001},
  \href{http://arxiv.org/abs/1905.09315}{{\ttfamily arXiv:1905.09315
  [hep-th]}}.

\bibitem{PhysRevD.54.7757}
A.~V. Smilga, ``Two-dimensional instantons with bosonization and physics of
  adjoint two-dimensional qcd,''
  \href{http://dx.doi.org/10.1103/PhysRevD.54.7757}{{\em Phys. Rev. D}
  {\bfseries 54} (Dec, 1996) 7757--7773}.
  \url{https://link.aps.org/doi/10.1103/PhysRevD.54.7757}.

\bibitem{PhysRevB.103.195113}
T.~Rudelius, N.~Seiberg, and S.-H. Shao, ``Fractons with twisted boundary
  conditions and their symmetries,''
  \href{http://dx.doi.org/10.1103/PhysRevB.103.195113}{{\em Phys. Rev. B}
  {\bfseries 103} (May, 2021) 195113}.
  \url{https://link.aps.org/doi/10.1103/PhysRevB.103.195113}.

\bibitem{doi:10.1063/5.0060808}
P.~Gorantla, H.~T. Lam, N.~Seiberg, and S.-H. Shao, ``A modified villain
  formulation of fractons and other exotic theories,''
  \href{http://dx.doi.org/10.1063/5.0060808}{{\em Journal of Mathematical
  Physics} {\bfseries 62} no.~10, (2021) 102301},
  \href{http://arxiv.org/abs/https://doi.org/10.1063/5.0060808}{{\ttfamily
  https://doi.org/10.1063/5.0060808}}. \url{https://doi.org/10.1063/5.0060808}.

\end{thebibliography}\endgroup


\end{document}